\documentclass[aps,pre,preprint,floatfix,superscriptaddress]{revtex4}
\usepackage[dvips]{graphicx}
\usepackage{latexsym}
\usepackage{natbib}
\usepackage[textwidth=3.7cm]{todonotes}
\usepackage{todonotes}
\usepackage{soul}
\usepackage{appendix}
\usepackage{amsmath, amssymb}
\usepackage{algorithm}
\usepackage{algpseudocode}
\usepackage{subfigure}
\usepackage{psfrag}

\begin{document}

\title{A study of cascading failures in real and synthetic power grid topologies using DC power flows}

\author{Russell Spiewak}
	\email[Corresponding author]{}
\author{Sergey V. Buldyrev}
\author{Yakir Forman}
\affiliation{Department of Physics,\\ Yeshiva University, \\500 West 185th Street, \\New York, NY 10033, USA\\Email: \{russell.spiewak,yakir.forman\}@mail.yu.edu, buldyrev@yu.edu}
\author{Saleh Soltan}
\author{Gil Zussman}
\affiliation{Department of Electrical Engineering,\\ Columbia University, \\1300 S.W. Mudd\\500 West 120th Street \\New York, NY 10027, USA\\Email: \{saleh,gil\}@ee.columbia.edu}

\date{\textbf{\today}}

\begin{abstract}

Using the linearized DC power flow model, we study cascading failures and their
spatial and temporal properties in the US Western
Interconnect (USWI) power grid. We  also introduce the preferential
Degree And Distance Attachment (DADA) model, with similar
degree distributions, resistances, and currents to the USWI.  We
investigate the behavior of both grids resulting from the
failure of a single line.  We find that the DADA model and the USWI
model react very similarly to that failure, and that their blackout
characteristics resemble each other.  In many cases, the failure of a single
line can cause cascading failures, which impact the entire grid.
We characterize the resilience of the grid by three parameters, the
most important of which is tolerance $\alpha$, which is the ratio of
the maximal load a line can carry to its initial load. 
We characterize a blackout by its yield, which we define as the
ratio of the final to the initial consumed currents. We find that
if $\alpha\geq2$, the probability of a large blackout occurring is very
small.  By contrast, in a broad range of $1<\alpha<2$, the initial
failure of a single line can result, with a high probability, in 
cascading failures leading to a
massive blackout with final yield less than 80\%.  The
yield has a bimodal distribution typical of a first-order
transition, i.e., the failure of a randomly selected line leads either
to an insignificant current reduction or to a major
blackout.  We find that there is a latent period in the
development of major blackouts during which few lines are overloaded, and
the yield remains high.  The duration of this latent period is
proportional to the tolerance.  The existence of the latent period
suggests that intervention during early time steps of a cascade can
significantly reduce the risk of a major blackout.

\end{abstract}

\maketitle

\section{Introduction}
\label{sec:intro}

The failure of a transmission line in a power grid leads to a
redistribution of power flows in the grid.  This redistribution may
cause overloads of other lines and their subsequent failures, leading
to a major blackout in a large area \cite{gil,Scala,Buldyrev,saleh}.
These failures may be initiated by natural disasters, such as
earthquakes, hurricanes, and solar flares, as well as by terrorist and
electromagnetic pulse (EMP) attacks \cite{EMP}.  Recent blackouts,
such as the 2003 and 2012 blackouts in the Northeastern
US and in India~\cite{2003,2012,India}, demonstrate that major power failure has a
devastating impact on many aspects of modern life.  Hence, there is a
dire need to study the properties of cascading failures in power grids.

The studies of failures in power grids usually employ the direct
current (DC) approximation of power flows
\cite{Gungor,saleh,bienstock1,bienstock2,pinar,Dobson,asz,Dobson2,Dobson3}.
In this paper, we employ a simplified DC model of a power grid which
is equivalent to a resistor network \cite{deArcangelis} and follow the
cascading failure model of \cite{saleh,gil}.
 
Ref.~\cite{Dobson} suggests that the distribution of blackouts in power grids follows a power law, which is related to the phenomenon of self-organized criticality. 
Other authors suggest that blackouts follow first-order phase transitions, in which the loss of power is either very small or very large \cite{Zapperi,Scala}.
The goal of this paper is to create a realistic model of a power grid which mimics the US Western Interconnect (USWI), to discover whether large blackouts can occur and to explore their distribution and spatial-temporal propagation, if they do occur. 
Additionally, we study the dependency of blackout characteristics on power grid design. 
For this reason, we introduce
a synthetic Degree And Distance Attachment (DADA) model \cite{manna,xulvi}.

We show that the characteristics of blackouts are universal. 
However, the sizes of blackouts are much smaller in the USWI-like model
with a realistic design than in an artificial DADA model with a
different spatial organization.  In particular, we study how the size
of the blackout and the dynamics of the cascading failures depend on
a set of three parameters of the model which characterize the
robustness of the grid: (1) tolerance $\alpha$ of
the lines to overload, compared to their initial loads
\cite{MotterLai,Motter}; (2) the minimum current $I_p$ which any line
in the network can carry independent of its initial load; and (3) the
interval of currents of the transmission lines $I_{u-\Delta u}$ to
$I_u$, from which a line is randomly selected to initiate the cascading
failures. We characterize $I_p$ and
$I_u$ by dimensionless parameters $p$ and $u$: the level of protection, $0<p\leq 1$, 
and the significance of initial failure, $0<u\leq 1$,
which are the fraction of the lines with transmitting
 currents less than $I_p$ and $I_u$, respectively. 

We find that in a broad range of $1 \leq \alpha < 2$, $u\geq 0.8$, and
$0<p<0.95$, large blackouts with final demand less than 80\% of the
initial demand may occur with a
significant probability both in the USWI-like grid and in artificially
constructed DADA grids.  Moreover, we find that in this range of
parameters the distribution of yield is bimodal, which is consistent with
first-order phase transitions.  Most importantly, we find that in each cascading failure which leads to a large blackout there is a latent
period during which the damage is localized in space, few lines fail,
and the decrease in yield is insignificant.  The existence of this
latent period suggests that the majority of blackouts can be
effectively stopped by the timely intervention of grid operators.  The
length of the latent period increases as the tolerance $\alpha$ increases.
Another important discovery is that in the event of a large blackout,
cascading failures stop when the network breaks into small,
disconnected islands.

The rest of the paper is organized as follows.  In Section
\ref{sec:models} we describe our methods of constructing power grids
and simulating cascading failures.
In Section \ref{sec:WestCoast} we study the properties of the USWI
model and its cascading failures. 
Next, in Section \ref{sec:prefModel} we describe the preferential
Degree And Distance Attachment (DADA) model. In Section
\ref{sec:comparison} we compare the physical features and behavior of
cascade simulations of the USWI model power grid and the DADA
model. 
Finally, in Section \ref{sec:conc} we discuss and summarize the
results of our study.

\section{Models}
\label{sec:models}
In this section, we describe the network, flow model, and cascade model in detail. 

\subsection{Definition of the Grid Elements}
\label{sec:elements}

We denote the set of all nodes as $N$ and the set of all lines as $E$.
Thus, a connected network is defined as the ordered pair $G=(N,E)$.
We denote the power grid network by the graph $G$ consisting of $n^0$
transmitting, $n^+$ supply, and $n^-$ demand nodes.  The total number
of nodes is $n=n^0+n^++n^-$.  Each supply or demand node is specified by the
amount of current it supplies ($I_i^+>0$) or by the amount of current
it demands ($I_i^->0$).  For transmitting nodes, we assume $I_i^0=0$.
Due to the law of charge conservation, $\sum I_i^+ = \sum I_i^-$.  The
network is specified by the $n\times n$ symmetric resistance matrix
with elements $R_{ij}$, where $R_{ij}$ is
the resistance equivalent to the reactance value of the transmission
line connecting nodes $i$ and $j$.  If there is no direct transmission
line connecting nodes $i$ and $j$, we assume that $R_{ij}=\infty$.
Since the matrix with elements $R_{ij}$ is
symmetric, the total number of transmitting lines $l$ is equal to the
number of finite elements of the resistance matrix divided by two.  Each
node $i$ is connected to $k_i$ nodes by transmission lines, where
$k_i$ is the number of finite elements in the $i^{th}$ row of the
resistance matrix. 
Thus $k_{i}$ represents the degree of node $i$ and $\left<k\right>$ is
the average degree of all nodes in the network. We denote the set
of all neighbors of node $i$ as $N(i)$.

\subsection{Flow Model}
\label{sec:flow}

We employ the simplified DC model of a power grid widely used in the
engineering community. 
This model is equivalent to flow equations in
resistor networks.  In this model the powers flowing through each line
are replaced by currents, the reactances of each line are represented
by resistances, and the phase angles are replaced by voltages.  Each
transmission line connecting nodes $i$ and $j$ is characterized by its
resistance $R_{ij}$, while each node
$i$ is characterized by its voltage $V_i$.  The current $I_{ij}$
flowing from node $i$ to node $j$ is
\begin{equation}
I_{ij}=(V_i-V_j)/R_{ij}.
\label{eq:OhmsLaw}
\end{equation} 
Additionally, the sum of all the currents flowing into each node $i$
is equal to the sum of all currents flowing out:
\begin{equation}
\sum_{j \in N(i)}I_{ij}=\delta^+I_i^+-\delta^-I_i^-,
\label{eq:JunctionRule}
\end{equation}
where $\delta_{i}^+=1$ or $\delta_{i}^-=1$ if a node $i$ is a supply
or a demand node respectively, and $\delta_i^+=\delta_i^-=0$
otherwise.  The particular methods used to solve the system of
equations (\ref{eq:OhmsLaw})-(\ref{eq:JunctionRule}) can be found in
Appendix \ref{sec:makematrix}.  

\subsection{Cascading Failures Model}
\label{sec:fails}

\begin{figure}[ht]
\centerline{
	\psfrag{AA}{\color[rgb]{1,0,0}$u$}
	\psfrag{BB}{\color[rgb]{1,0,0}$u-\Delta u$}
	\psfrag{PP}{\color[rgb]{0,0.56,0}$p$}
	\psfrag{II}{\color[rgb]{0,0,1}\,\,$I_1$}
	\psfrag{AII}{\color[rgb]{0,0,1}\,\,$\alpha I_1$}
	\psfrag{IP}{\color[rgb]{0,0.56,0}$I_p$}
	\psfrag{IIS}{\color[rgb]{0,0,1}\,$=I_1^*$}
	\psfrag{III}{\color[rgb]{0,0,1}\,\,$I_2$}
	\psfrag{AIII}{\color[rgb]{0,0,1}$\alpha I_2$}
	\psfrag{IIIS}{\color[rgb]{0,0,1}\,\,$=I_2^*$}
	\psfrag{IAA}{\color[rgb]{1,0,0}$I_u$}
	\psfrag{IBB}{\color[rgb]{1,0,0}$I_{u-\Delta u}$}
	\psfrag{LL}{\color[rgb]{1,0,0}Initial Failure}
	\includegraphics[clip, width=0.8\linewidth,angle=0]{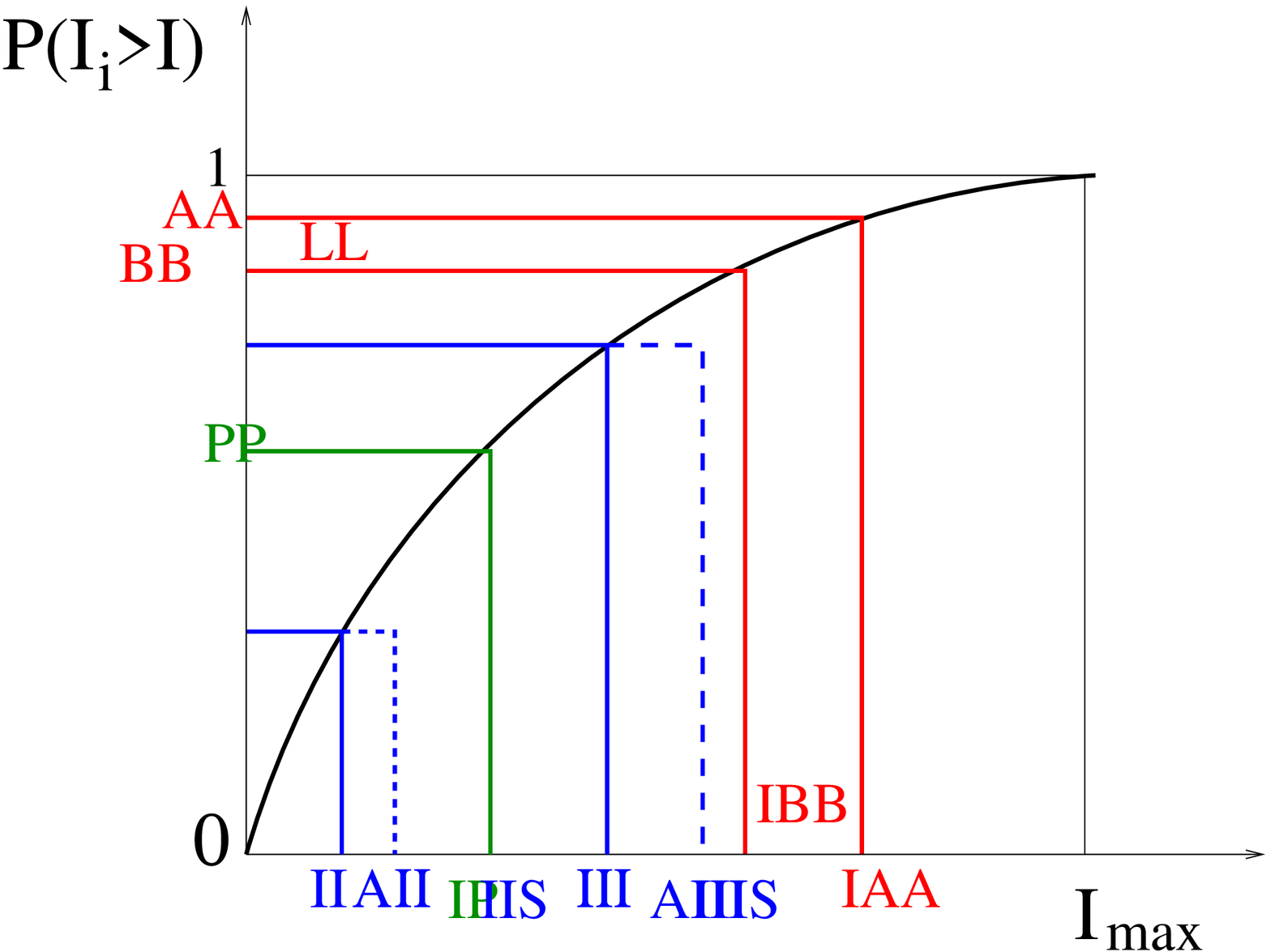}}
\caption{(Color online) Schematic illustration of the rules of cascading failures. 
The black solid line represents the cumulative distribution of currents in transmission lines. 
The green solid line indicates the fraction $p$ of uniformly protected lines and the current $I_p$ corresponding to this fraction. 
Blue solid lines indicate the initial loads of two transmission lines $I_1$ and $I_2$.
Blue dashed lines indicate the maximum loads of these lines $\alpha I_1$ and $\alpha I_2$, defined using the tolerance parameter $\alpha$. 
Since $\alpha I_1<I_p$, the actual maximum load for line 1 is $I_p$.  
In contrast, since $\alpha I_2>I_p$, the actual maximum load for line 2 is $\alpha I_2$. 
Red lines show the maximum and the minimum fraction of lines, and their corresponding currents, from which region the initially failed lines can be selected for given parameters $u$ and $\Delta u$.}
\label{f:rules}
\end{figure}

Once the system (\ref{eq:OhmsLaw})-(\ref{eq:JunctionRule}) is solved,
we find the currents in all transmitting lines $I_{ij}$ and define
their maximum capacities $I_{ij}^\ast$ using the following two rules
(Fig.~\ref{f:rules}): (i) we define $I_p$ as the standard capacity of
the lines.  It is computed such that a fraction $p$ of the lines
initially have currents below $I_p$. We refer to $p$ as the {\it level
  of protection}.  (ii) For each line we define its individual
capacity $\alpha |I_{ij}|$, where $\alpha\geq 1$ is the {\it tolerance}
(i.e., the factor of safety).  We assume $\alpha$ to be the same for
every transmission line in the grid.  Using these rules,
\begin{equation}
I_{ij}^\ast \equiv {\rm max}(I_p,\alpha |I_{ij}|).
\end{equation}
If a current in line $\{ij\}$ exceeds $I^\ast_{ij}$, the line will fail.

The larger $p$ and $\alpha$ are, the better the grid is protected
against overloads.  We usually use $p=0.9$ and vary $\alpha$ as the
main parameter of grid resilience.  We introduce the standard
capacity $I_{p}$ because it would be unrealistic to assume
that lines are built with very low capacity.  In fact, many lines in
realistic power grids may be built for backup reasons, so initially
$I_{ij}=0$.  Without standard load requirements any redistribution of
currents in the grid would lead to the failure of
these backup lines, which contradicts their purpose. 

\begin{figure}
\begin{algorithm}[H]
\caption{Cascading Failures}
\label{alg:cascs}
\begin{algorithmic}[1]
	\Require $G_0=\left(N,E\right)$, $\alpha$, $p$, $u$.
        \State $E_0 \gets E$.
	\State  $\forall \{i,j\}\in E$ calculate $I_{ij}$.
	\State Determine $I_p$ for given $p$.
	\State Choose a line $\{i,j\}$ from the interval $[I_{u-\Delta u},I_u]$ of currents to fail; $s\gets 0$, $F_0 \gets \left\{i,j\right\}$ ($s$ is stage of cascade; $F_s$ is set of lines failed during $s^{th}$ stage of the cascade). 
	\While{$F_{s}\neq \emptyset$}
		\For {each cluster $C$}\label{alg:clustPow}
			\State Calculate $\sum_{i\in C} I_i^{+(s-1)}$ and $\sum_{i \in C} I_i^{-(s-1)}$.
			\If {$\sum_{i \in C} I_i^{+(s-1)}>\sum_{i\in C} I_i^{-(s-1)}$}
				\State $\forall i$: $I_i^{+(s)} \gets I_i^{+(s-1)}*\left(\frac{\sum_{i\in C} I_i^{-(s-1)}}{\sum_{i\in C} I_i^{+(s-1)}}\right) $ 
			\ElsIf {$\sum_i I_i^{-(s)}>\sum_i I_i^{+(s)}$}
				\State $\forall i$: $I_i^{-(s)} \gets I_i^{-(s-1)}*\left(\frac{\sum_{i\in C} I_i^{+(s-1)}}{\sum_{i\in C} I_i^{-(s-1)}}\right)$
			\EndIf
		\EndFor	
		\State $E_{s+1} \gets E_s\setminus F_s$.
		\State $G_{s+1} \gets G_{s}$ ($G_s$ is the state of the grid at the $s^{th}$ stage of the cascade).
		\ForAll{$\{i,j\}\in E_s$}
			\State Calculate $I_{ij}^{(s)}$. 
			\If {$|I_{ij}^{(s)}|>I_{ij}^\ast$}
				\State $F_{s+1} \gets F_{s+1}\bigcup\left\{i,j\right\}$.
				\State $G_{s+1} \gets G_{s+1}\setminus\left\{i,j\right\}$.
			\EndIf	
		\EndFor	
		\State $s \gets s+1$.
	\EndWhile
\end{algorithmic}
\end{algorithm}
\label{a:failure}
\end{figure}

To initiate a cascading failure, we randomly select and remove a
single line for which the current $|I_{ij}|$ belongs to the interval
of currents $[I_{u-\Delta u},I_u]$, where the fraction $u-\Delta u$ of
lines operate below $I_{u-\Delta u}$ and the fraction $u$ of lines
operate below $I_u$ (Fig.~\ref{f:rules}).  Throughout this paper we
select $\Delta u=0.1$.  Precisely, the parameter $u$ specifies the
significance of the lines which are targeted for the initial failure.
We refer to $u$ as the {\it significance of initial failure}.  For
example, $u=1.0$ and $\Delta u=0.1$ means that the lines which
initially fail are selected from the top 10\% of lines ranked
according to their initial current. 

Real power grids are usually designed in such a way that the removal
of a single element does not cause any instability in the system.
This condition is called the $N-1$ property.  For simplicity, this
model does not have the $N-1$ property.  However, removing a single
line from a network without the $N-1$ property is the equivalent to
removing two lines from a network with the $N-1$ property. 
Therefore, the properties of blackouts will be the same for networks without the
$N-1$ property as for networks with the $N-1$ property, given equivalent line removals. 
The only difference is that the network with the
$N-1$ property will be more stable because the probability of
simultaneous failure of two significant lines is much smaller than the
probability of failure of one significant line.

It should be pointed out that removing a line can lead to
disintegration of the grid into two disconnected components, which we
call clusters.  Obviously, the supply and demand in each cluster
should be equalized to retain charge conservation. 
Thus, for each cluster $C_j$, we compute $\sum_{i\in C_j} I_i^+$ and
$\sum_{i\in C_j}I_i^-$.  If in a cluster $\sum_{i\in C_j} I_i^+ >
\sum_{i\in C_j} I_i^-$, we multiply the current of each supply node in
$C_j$ by $\sum_{i\in C_j} I_i^- / \sum_{i\in C_j} I_i^+ <1$; if
$\sum_{i\in C_j} I_i^- > \sum_{i\in C_j} I_i^+$, we multiply the
current of each demand node in $C_j$ by $\sum_{i\in C_j} I_i^+ /
\sum_{i\in C_j} I_i^-<1$ (see Algorithm
\algref{alg:cascs}{alg:clustPow}).  In this equalization method, we spread
the decrease in current uniformly among all the supply or demand nodes
in $C_j$.

Then, we modify 
Eqs.~(\ref{eq:OhmsLaw})-(\ref{eq:JunctionRule}), solve the
resulting system of equations, and find new potentials $ V^{(1)}_i$ and
new currents $I^{(1)}_{ij}$.  We also compute
the total number of surviving lines $l_1=l-1$ and the total supplied
current $I_1=\sum_i I_i^{+(1)}=\sum_i I_i^{-(1)}$, where $I_i^{+(1)}$
and $I_i^{-(1)}$ are the new supply and demand currents, computed as
described above if clusterization has occurred.  We define this
situation as the first time step $t=1$ of the cascade of failures.

At the second time step of the cascade, we remove all lines for which
the new current $|I^{(1)}_{ij}|$ exceeds the predefined maximum load
of this line, $I^\ast_{ij}$.  If no overloads have occurred, the
cascade has stopped and we assume $I_f=I_1$ to be the final demand
current of the process.  If some of the lines fail, we repeat
the equalization algorithm, modify the
system of equations
(\ref{eq:OhmsLaw})-(\ref{eq:JunctionRule}),
calculate $ V^{(2)}_i$ and new currents $I^{(2)}_{ij}$, and compute the new
total demand current $I_2$ and the new total number of active lines
$l_2$.

We repeat this process recurrently until at a certain time step $t$ of
the cascade no lines fail.  We call this time step the final time step
of the cascade, and compute $I_f=I_t$, $l_f=l_t$, and the duration of
the cascade $f=t$.  The cascading failures algorithm is summarized in
Algorithm \ref{alg:cascs}.

\begin{table*}[ht]
\begin{tabular}{|c|p{0.9\linewidth}|}
\hline
\multicolumn{2}{|l|}{{\bf Table 1.} Parameters of the model}\\\hline
{\bf $\boldsymbol{n^+}$} & The number of supply nodes\tabularnewline\hline 
{\bf $\boldsymbol{n^-}$} & The number of demand nodes\tabularnewline\hline 
{\bf $\boldsymbol{n^0}$} & The number of transmitting nodes\tabularnewline\hline 
{\bf $\boldsymbol{I_i^+}$} & The current supplied by supply node $i$\tabularnewline\hline 
{\bf $\boldsymbol{I^-_i}$} & The current demanded by demand node $i$\tabularnewline\hline
{\bf $\boldsymbol{R_{ij}}$} & The resistance of the line connecting nodes $i$ and $j$\tabularnewline\hline
{\bf $\boldsymbol{V_i}$} & The voltage of node $i$\tabularnewline\hline
{\bf $\boldsymbol{I_{ij}}$} & The current traveling through the line connecting nodes $i$ and $j$\tabularnewline\hline
{\bf $\boldsymbol{\alpha}$} & The tolerance of the lines\tabularnewline\hline
{\bf $\boldsymbol{p}$} & The level of protection of the lines\tabularnewline\hline
{\bf $\boldsymbol{u}$} & The significance of the initial failure\tabularnewline\hline
\end{tabular}
\end{table*}

\subsection{Metrics}
\label{sec:metrics}

We define here all the metrics we use to characterize the cascades of failures.\\
{\bf Cascade Duration, $\boldsymbol{f}$}: the number of time steps until a cascade stops.\\ 
{\bf Number of Active Lines, $\boldsymbol L=l_f$}: the number of transmission lines in the grid that have not failed by the end of the cascade.\\
{\bf Yield, $\boldsymbol{Y(t)}$}: $\frac{I_t}{I_0}$, the ratio between the demand at time step $t$ ($I_t$)
and the original demand ($I_0$). 
For $t=f$, we simply denote yield by $Y$.\\
{\bf Large Blackout}: $Y\leq 0.8$, when the yield in the grid drops below the threshold of 80\%.\\
{\bf Risk of Large Blackout, $\boldsymbol{\Pi(\alpha)}$}: the probability that the failure of a line will produce a Large Blackout with $Y<0.8$.\\
{\bf $\boldsymbol{\left<G\right>}$}: the average relative size (fraction of nodes) of the largest cluster at the end of the cascade. \\
{\bf $\boldsymbol{\left<Y\right>}$}: the average yield at the end of the cascade.\\
{\bf $\boldsymbol{\left<L\right>}$}: the average number of surviving lines at the end of the cascade.\\
{\bf Hop distance, $\boldsymbol{h_i}$}: the shortest path measured in number of lines required to reach a certain node $i$ from the failed line.\\
{\bf Local Yield, $\boldsymbol{Y(t,h)}$}: 
\begin{equation}
Y(t,h) = \frac{\sum_{i\in H(h)} I^{-(t)}_i}{\sum_{i\in H(h)} I^-_i}
\end{equation}
where $H(h)$ is the subset of demand nodes a given hop distance $h$
from the failed line.\\ {\bf Blackout Radius of Gyration},
$\boldsymbol{r_B}(t)$: a quantitative measure of the blackout's geometric
dimension as a function of the cascade time step $t$,
\begin{equation}
r_B(t)^2 = \frac{\left\langle\sum_{i\in B(t)} h_i^2  I^{-(t)}_i\right\rangle}{\left\langle\sum_{i\in B(t)} I^-_i\right\rangle},
\end{equation}
where the summation is made over the set $B(t)$ of totally disconnected demand nodes which do not receive
any current at the $t^{th}$ time step of the cascade. The average is done either over all runs resulting in a large blackout or all runs not resulting in a large blackout.

{$\boldsymbol{t_l}$}: duration of the latent period, i.e., the time step of the
cascade of failures at which $Y(t_l)$ drops below 95\%.\\

\section{Cascades in USWI}
\label{sec:WestCoast}

In this section we simulate the cascades on the USWI network obtained
from \cite{gil}.  The network is based on real power grid topology
data of the western US taken from the Platts Geographic Information
System (GIS) \cite{gis} and includes approximate information about
transmission lines based on their lengths, supplies based on power plants'
capacities, and demands based on
the population at each location.

\subsection{USWI Properties}
\label{sec:USWIdata}

The USWI power grid data  contains 8050  transmitting
 nodes, 1197 supply nodes,
3888 demand nodes, and 17544 
transmission lines connecting them.  To avoid exposing possible
vulnerabilities in the actual USWI, our data set does not include the
geographic coordinates of nodes.  It does, however, include the
length of each line $r_{ij}$ connecting nodes $i$ and $j$.  We define
the resistances of the lines to be proportional to their lengths
$R_{ij}=\rho r_{ij}$, where $\rho$ is a constant.  The data set also
specifies the supply ($I_i^+$) and demand ($I_i^-$) of each relevant
node.

\subsubsection{Degree Distribution}
\label{sec:USWIDegDist}

\begin{figure*}[ht]
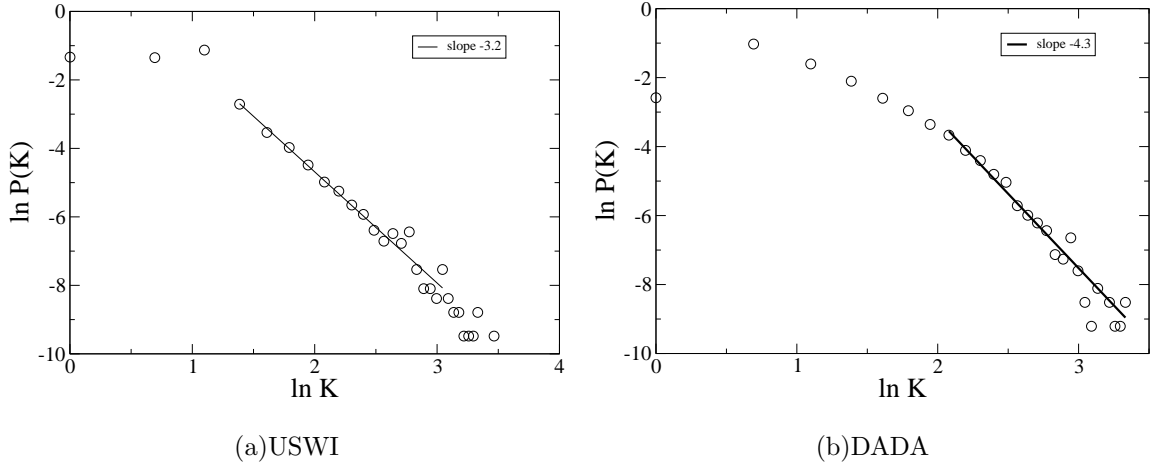

\centering
  \subfigure[USWI]{
	  \includegraphics[clip, width=0.45\linewidth,angle=0]{Fig2a.eps}
	  \label{subfig:DegUSWI}
  }
  \subfigure[DADA]{
	  \includegraphics[clip, width=0.45\linewidth,angle=0]{Fig2b.eps}
	  \label{subfig:DegDADA}
  }
\caption{Degree Distributions of \subref{subfig:DegUSWI}: the USWI power grid, and \subref{subfig:DegDADA}: the DADA model for $\mu=6$, $\ell=1.5$.}
\label{f:deg}
\end{figure*}

The USWI power grid is characterized by a fat-tail degree distribution
of nodes (Fig.~\ref{subfig:DegUSWI}), which can be approximated by a
power law $P(k)\approx k^{-3}$ with an exponential cut-off.  The degree
distributions of transmitting nodes, supply nodes, and demand nodes are
quite similar to each other (Fig.~\ref{subfig:DegUSWI}).  The average
degree $\left< k\right>$ of the nodes in the USWI is 2.67.  For supply
nodes, it is slightly larger (2.88); for demand nodes, it is very
slightly smaller (2.61).

\subsubsection{Length Distribution}
\label{sec:USWILengthDist}

The length distribution of lines for the USWI has an approximately lognormal shape with
power law tails.  Figure~\ref{subfig:ResUSWI} shows $\ln P(\ln
r_{ij})$, the logarithm of the probability density function (PDF) of
$\ln r_{ij}$.  For the lognormal distribution the graph would be a
perfect parabola.  Instead, we see that both tails of the distribution
can be well approximated by straight lines with slope $\nu_-=0.77$ for
the left tail and slope $\nu_+=-1.44$ for the right tail.  This means
that the PDF of $r_{ij}$  can be approximated
by power laws $P(r)\approx r^{\nu_--1}$ for $r\to 0$ and $P(r)\approx
r^{\nu_+-1}$ for $r\to\infty$.  The extra term $-1$ comes from the
equivalence of cumulative distribution functions $P(r>r_0)=P(\ln r>\ln
r_0)$, from which, differentiating with respect to $r_0$, we obtain
$P(r_0)=P(\ln r_0){\mathrm d}\ln(r_0)/{\mathrm d}r_0=P(\ln(r_0))/r_0$.

A very small power for the left tail indicates an intriguing
possibility that the USWI forms a fractal set with fractal dimension
$D=0.77<2$. Indeed, if the mass (number of nodes) within a circle of radius $r$ scales as
$a r^D$, it follows from the Poisson
distribution that the probability to find an empty circle of radius
$r$ surrounding a given power station is $\exp(-a r^D)$.  The derivative of this function is
$\sim r^{D-1}$ for $r\to 0$.  This is the probability
density of the distances to the nearest neighbors, which should be a
good approximation to the left tail of $P(r)$.  Accordingly, the slope of the left tail 
$\nu_-=D$.  The fractality of the population
distribution was suggested in a number of works \cite{Batty,Makse}.
This is particularly plausible for the US western regions, in which
densely populated areas are separated by deserts and mountains.

\subsubsection{Distribution of Supply, Demand, and Line Currents}
\label{sec:USWIPowers}

The supplies (demands) in the USWI approximately follow a lognormal
distribution, with a sharp cut-off in the right tail, indicating the
existence of a technical upper limit for the supplies (demands) (see
Fig.~\ref{f:pow}).  Using this data we solve equations
(\ref{eq:OhmsLaw})-(\ref{eq:JunctionRule}) and find the currents of
transmission lines $I_{ij}$.  The cumulative distribution function of
currents in the USWI model is
roughly exponential (Fig.~\ref{f:curr}), with about 10\% extremely
small currents.

\subsection{Properties of Cascades in USWI}
\label{sec:yield}
\subsubsection{Bimodality of the Yield Distribution}
\label{sec:Bimodality}
The most important metric for studying
cascading failures is the yield $\boldsymbol Y$.  The yield
critically depends on the initially failed line.  We fix parameters of
the model ($\alpha$ and $p$) and simulate the cascade initiated by the
failure of a line selected from an interval of currents $[I_{u-\Delta
    u},I_u]$.  We measure yield and repeat this procedure $Q=100$
times.  Next we construct the histogram of yields.

\begin{figure*}[h]
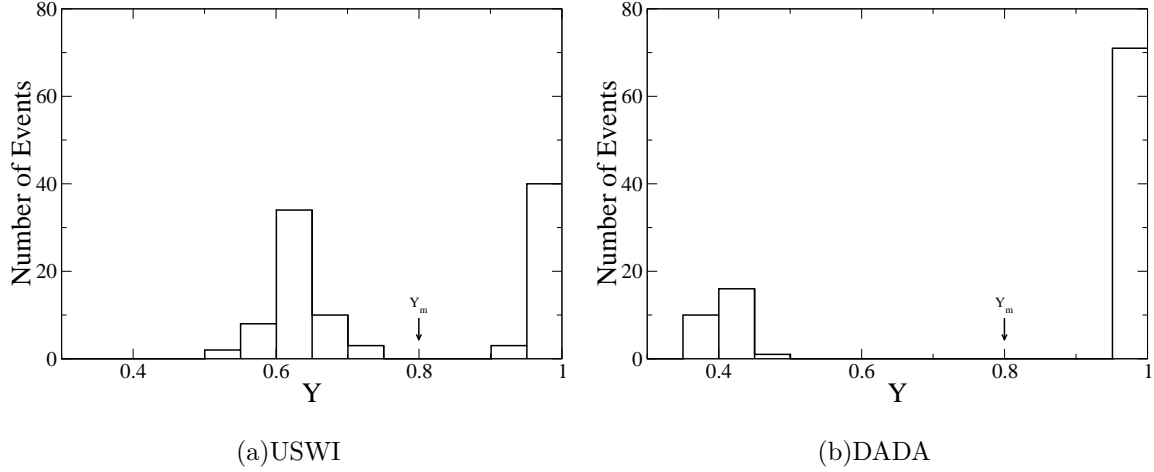

\centerline{
  	\subfigure[USWI]{	
  		\includegraphics[clip, width=0.45\linewidth,angle=0]{Fig6a.eps}
  		\label{subfig:HistUSWI}
  }
  \subfigure[DADA]{
  		\includegraphics[clip, width=0.45\linewidth,angle=0]{Fig6b.eps}
  		\label{subfig:HistDADA}
  }}
\caption{Distribution of yield for $\alpha=1.6$, $p=0.9$, and
$u=1.0$. One can clearly see the bimodality of the distribution with two peaks
for high yield 0.975 and low yield 0.625, with practically no yields between
0.75 and 0.9 for the USWI \subref{subfig:HistUSWI}. Similarly for the DADA, one can clearly see the bimodality of the distribution with two peaks for high yield 0.975 and low yield 0.425, with practically no yields between 0.5 and 0.95 \subref{subfig:HistDADA}.}
\label{f:hist}
\end{figure*}

\begin{figure*}[h]
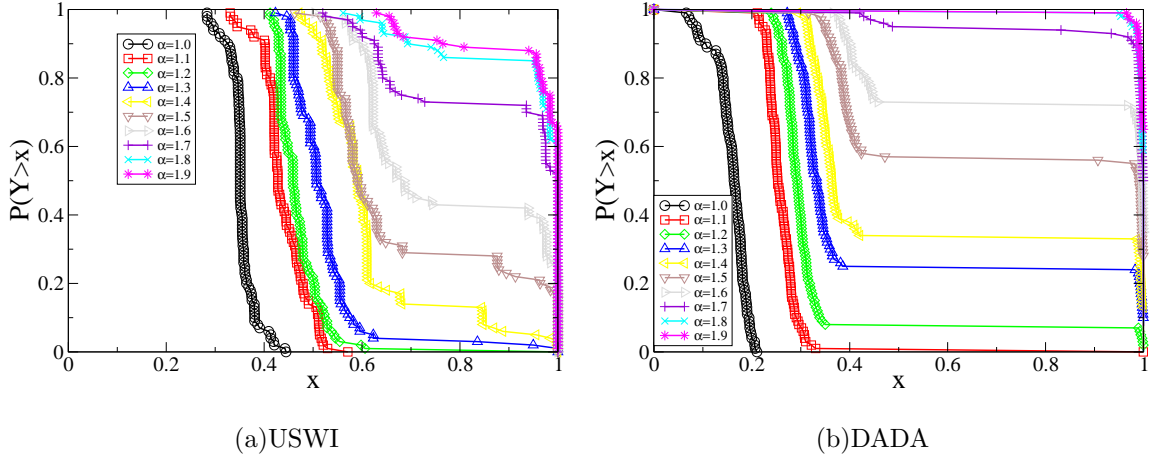

\centering
  \subfigure[USWI]{
	  \includegraphics[clip, width=0.45\linewidth,angle=0]{Fig7a.eps}
	  \label{subfig:PifUSWI}
  }
  \subfigure[DADA]{
	  \includegraphics[clip, width=0.45\linewidth,angle=0]{Fig7b.eps}
	  \label{subfig:PifDADA}
  }
\caption{Cumulative distribution of the yield for $p=0.9$, $u=1.0$ and various values of $\alpha$ for \subref{subfig:PifUSWI} the USWI model, and \subref{subfig:PifDADA} the DADA model with $\mu=6$, $\ell=1.5$. The large gap in the distributions is a feature of the abrupt first-order transition. For the USWI, $u=1.0$ means that the distribution is obtained by running 100 simulations with the initial removal of one of the top 100 lines with the largest initial currents.}
\label{f:PIf}
\end{figure*}

The interesting feature of the yield histogram is its pronounced
bimodality (Fig.~\ref{subfig:HistUSWI}), which can be detected by a
plateau in the cumulative yield distribution
(Fig.~\ref{subfig:PifUSWI}).  The bimodality of the yield distribution
is present in a large region of the parameter space ($\alpha,p,u$),
characterized by relatively small $\alpha<2$, practically all $p\leq
0.95$, and relatively large $u>0.8$.  The tolerance parameter $\alpha$
is identical to the tolerance $\alpha+1$ of the Motter model
\cite{MotterLai,Motter}.  We find the same general behavior as in the
Motter model, that as $\alpha$ increases the probability of a large
blackout decreases (Fig.~\ref{subfig:PifUSWI}).

One can see (Fig.~\ref{subfig:PifUSWI}) that the distribution of yield
clearly remains bimodal for $\alpha<2$ for the USWI model.  Similar
behavior can be observed in the Motter model and the mutual percolation
model, where the collapse transition is shown to be a first-order transition
(all-or-nothing transition)
\cite{Zapperi,MotterLai,Motter,Buldyrev,Kornbluth}.  In Section
\ref{sec:metrics} we defined large blackouts as having $Y<0.8$.  The
motivation behind this definition is that there is a gap in the
cumulative distribution of the yield between $Y=Y_1$ and $Y=Y_2$ that
separates severe cascades from mild ones.  It is possible to select
$Y_m=0.8$ as a value of $Y$ that belongs to the interval
$[Y_1(\alpha,p,u), Y_2(\alpha,p,u)]$ for a large section of the
parameter space where the bimodality is observed. Thus this value can
be used as a universal threshold which separates severe blackouts from
mild blackouts.  For large $\alpha$, the gap in the yield distribution
reduces and eventually disappears for $\alpha\geq2$.

\subsubsection{Risk of Large Blackouts}
\label{sec:riskBlackouts}

The ensemble of cascades can be characterized by two important
parameters of the outcome: (i) the probability of a large blackout
$P(Y<0.8)$, which we call the risk of a large blackout $\Pi(\alpha)$,
and (ii) the average blackout yield $\left<Y\right>$, for the cases of
large blackouts.

\begin{figure*}[h]
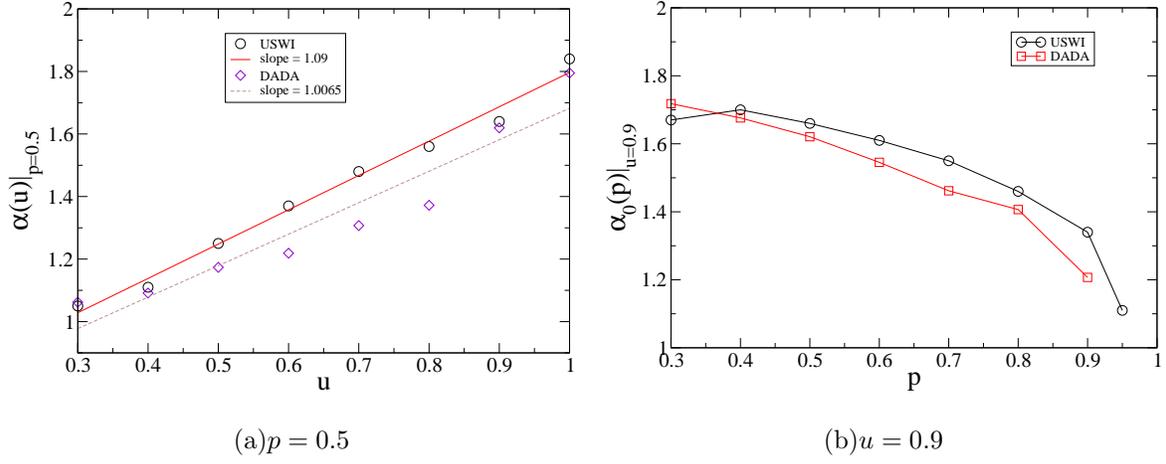

\centerline{
  \subfigure[$p=0.5$]{
  		\includegraphics[clip, width=0.46\linewidth,angle=0]{Fig9a.eps}
  		\label{subfig:Alpha0-p0.5}
  }
  \subfigure[$u=0.9$]{
  		\includegraphics[clip, width=0.45\linewidth,angle=0]{Fig9b.eps}
  		\label{subfig:Alpha0-u0.9}
  }}
\caption{ Behavior of $\alpha_0(u,p)$ defined as the value of $\alpha$
  such that $r(\alpha,u,p)=0.5$ as function of $u$ at constant $p=0.5$
  \subref{subfig:Alpha0-p0.5} and as function of $p$ at constant
  $u=0.9$ \subref{subfig:Alpha0-u0.9}.}
\label{f:alpha0}
\end{figure*}

Figs.~\ref{subfig:YvA-p0.5-USWI} and \ref{subfig:YvA-u0.9-USWI} show
how the risk of large blackouts $\Pi(\alpha)$ decreases as $\alpha$
increases for different values of $u$ and $p$.  We find that for
different values of $u$ and $p$, the shapes of the curves $\Pi(\alpha)$ remain
approximately constant, but the curves significantly shift in a
horizontal direction.  This means that the curves $\Pi(\alpha)$ can be
well approximated by $\Pi\left(\alpha -\alpha_0(u,p)\right)$.  The
function $\alpha_0(u,p)$ can be defined by solving the equation
$\Pi\left(\alpha_0(u,p)\right)=1/2$ with respect to $\alpha_0(u,p)$.
One can see that $\alpha_0(u,p=0.5)$ is an approximately linear
function of $u$, which increases with $u$
(Fig.~\ref{subfig:Alpha0-p0.5}).  This means that the higher the
current of the initially failed line, the larger the tolerance
necessary to achieve the same degree of protection for the
transmission lines.  In other words, the same effect can be achieved
either by protecting a certain fraction of the most significant lines
from spontaneous failure, or by
increasing the tolerance of all the lines by some quantity
(Fig.~\ref{subfig:Alpha0-p0.5}).  The dependence of the risk on $p$ is
weaker than on $u$, especially for $p\leq 0.5$
(Fig.~\ref{subfig:Alpha0-u0.9}).  An increase in $p$ has practically
no effect on increasing the robustness of the grid.  The increase in
$p$ achieves a significant effect on the risk of large blackouts only
when $p$ approaches $0.9$.

\subsubsection{Characteristics of Large Blackouts}
\label{sec:LargeBlacks}

\begin{figure*}[h]
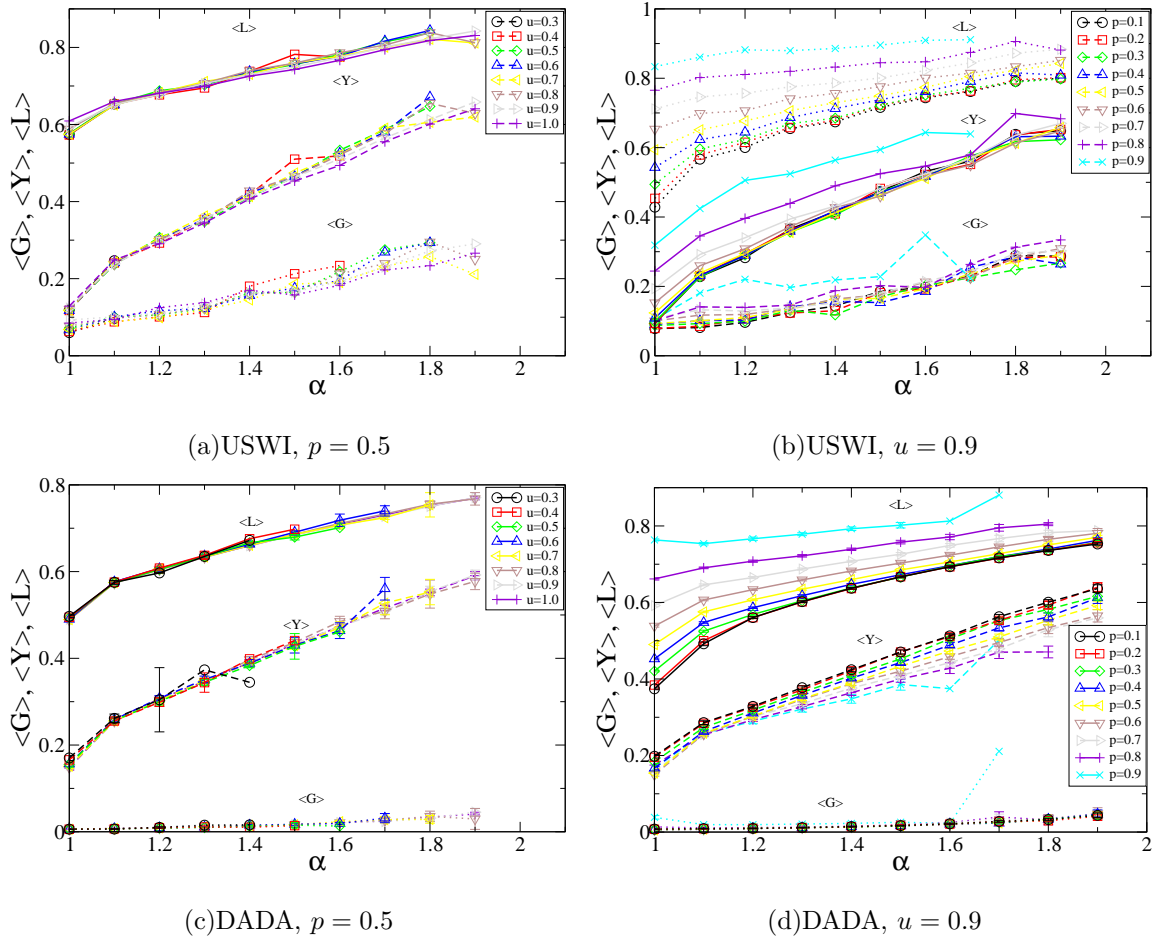

\centerline{
  \subfigure[USWI, $p=0.5$]{
  		\includegraphics[clip, width=0.45\linewidth,angle=0]{Fig10a.eps}
  		\label{subfig:b-USWI-u}
  	}
  \subfigure[USWI, $u=0.9$]{
  		\includegraphics[clip, width=0.45\linewidth,angle=0]{Fig10b.eps}
  		\label{subfig:b-USWI-p}
  }} 
\centerline{
  \subfigure[DADA, $p=0.5$]{
  		\includegraphics[clip, width=0.45\linewidth,angle=0]{Fig10c.eps}
  		\label{subfig:b-DADA-u}
  }
  \subfigure[DADA, $u=0.9$]{
  		\includegraphics[clip, width=0.45\linewidth,angle=0]{Fig10d.eps}
  		\label{subfig:b-DADA-p}
  }}
\caption{ Behavior of the yield $\langle Y\rangle$, the fraction of nodes in the largest connected component $\langle G\rangle$, and the fraction of survived lines $\langle L \rangle$ averaged only over runs which resulted in large blackouts ($y<0.8$) as a function of $\alpha$ for different $u$ and a fixed value of $p=0.5$ \subref{subfig:b-USWI-u} USWI model and \subref{subfig:b-DADA-u} DADA model, and different $p$ at fixed $u=0.9$ \subref{subfig:b-USWI-p} USWI model and \subref{subfig:b-DADA-p} DADA model. 
The small $\left<G\right>$ in the DADA model is a feature of the symmetry of the model network and the long lengths of the first links constructed (as necessitated by \cite{asz}).}
\label{f:GYL}
\end{figure*}

Large blackouts can be characterized by their average yield $\langle
Y\rangle$, average fraction of surviving lines $\langle L \rangle$,
and the average fraction of nodes in the largest connected component
of the grid $\langle G\rangle$.  These metrics only weakly depend on
$u$, but are strongly increasing functions of $\alpha$
(Fig.~\ref{subfig:b-USWI-u}).  The independence of the characteristics
of large blackouts on $u$ stems from the fact that the properties of
large blackouts, if they occur, do not depend on a particular line to
initiate the failure.  The risk of large blackouts depends on $u$, but
the average parameters of large blackouts do not.

The dependence of these metrics on $p$ is more complex
(Fig.~\ref{subfig:b-USWI-p}).  While the yield $\langle Y\rangle$
starts to increase only for $p>0.7$, the number of survived lines
$\langle L\rangle$ significantly increases with $p$ even for small
$p$.  This is not surprising since $p$ is the level of protection of
the lines, and
fewer lines fail if more lines are protected.  As $p$ approaches $1$,
the dependence of $\langle L\rangle$ on $\alpha$ becomes very weak. The
explanation of this fact is based on the notion that $\langle
L\rangle$ is computed only
for the case of {\it large} blackouts. 
 For a large blackout to occur, a significant fraction of
lines must fail, sufficient to disconnect a large fraction of demand
nodes.  On the other hand, as $\alpha$ increases, the risk of a large
blackout decreases to zero, so the average fraction of lines
 surviving for all the cascades (large
and small) approaches 1.

Another important observation from Fig.~\ref{subfig:b-USWI-p} is the
very small dependence of $\langle G\rangle$ on the parameters
$\alpha$, $p$, and $u$, as opposed to $\langle L\rangle$.  Hence, by
removing a small fraction of the lines (20\%) the grid disintegrates
into many small clusters, each less than 20\% of the total size.
Indeed, percolation theory predicts that close to the percolation
threshold, it is sufficient to delete an infinitesimally small
fraction of the so called ``red'' bonds (which form a fractal set with
fractal dimension $3/4$) to divide the network into a set of
small disconnected components \cite{Coniglio}.

\subsection{Latent Period of the Cascade}
\label{sec:cascs}

The cascading failures that do not result in large blackouts
($Y>0.8$) are usually short $(f< 8)$
(Figs.~\ref{subfig:tVsAlph-USWI-p0.5} and
\ref{subfig:tVsAlph-USWI-u0.9}).  In contrast, the duration of
cascades resulting in large blackouts $Y\leq 0.8$ increases with
$\alpha$, reaching values of order 40 for large $\alpha$.  This means
that for large tolerances, it takes much longer for the blackout to
spread over a large area, since at each time step only a few lines have
a huge overload and fail.

\begin{figure*}[h]
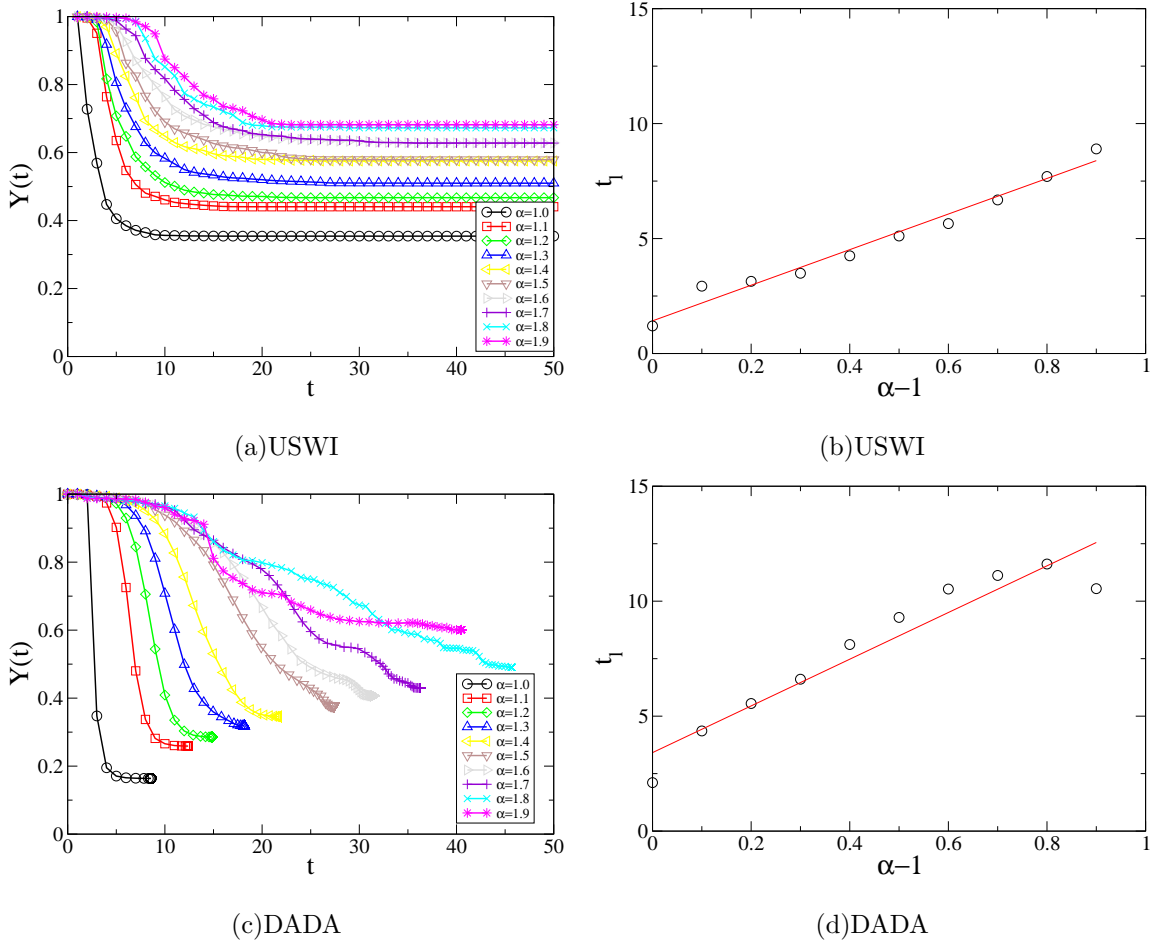

\centerline{
  \subfigure[USWI]{
  		\includegraphics[clip, width=0.45\linewidth,angle=0]{Fig12a.eps}
  		\label{subfig:Y-USWI}
  }
  \subfigure[USWI]{
  		\includegraphics[clip, width=0.45\linewidth,angle=0]{Fig12b.eps}
  		\label{subfig:Alph-1-USWI}
  }}
\centerline{
  \subfigure[DADA]{
  		\includegraphics[clip, width=0.45\linewidth,angle=0]{Fig12c.eps}
  		\label{subfig:Y-DADA}
  }
  \subfigure[DADA]{
  		\includegraphics[clip, width=0.45\linewidth,angle=0]{Fig12d.eps}
  		\label{subfig:Alph-1-DADA}
  }}  
\caption{ Decrease in the demanded current as a
  function of the cascade time step during the cascades resulting in
  large blackouts for \subref{subfig:Y-USWI} the USWI model and
  \subref{subfig:Y-DADA} the DADA model.  The latent period at the
  beginning of the cascade, during which there is no significant
  decrease in the demand, linearly increases with tolerance
  $\alpha$ for both \subref{subfig:Alph-1-USWI} the USWI model and
  \subref{subfig:Alph-1-DADA} the DADA model.  In both USWI and DADA
  models $u=1.0$, $p=0.9$.}
\label{f:latent}
\end{figure*}

In the cascades resulting in large blackouts, the fraction of the
consumed current, $Y(t)$, decreases with time in a non-trivial way
(Fig.~\ref{subfig:Y-USWI}). During the first few time steps of the
cascades, the yield does not significantly decrease since the current
can successfully redistribute over the remaining lines without
disconnection of the demand nodes.  This period, in which the cascade
is still localized and a blackout has not yet occurred, can be called
the {\it latent period of the cascade}. The recognition of this
latent period is important since it is a period in which a cascade is
beginning to spread but has not yet grown uncontrollable. In the
latent period it may still be possible to intervene and 
redistribute current flow to stop the cascade before it becomes a
large blackout.

We define the duration of this latent period of the cascade as the
time step at which the yield drops below 0.95. At approximately this
time step the yield starts to rapidly decrease and then, towards the
end of the cascade, stabilizes again.  The shape of this function is
characteristic of an abrupt first-order transition observed in simpler
models of network failure \cite{Buldyrev,Motter}.  Remarkably, the
duration of the latent period is a linear function of tolerance
(Fig.~\ref{subfig:Alph-1-USWI}).

\subsection{Cascade Spatial Spreading}
\label{sec:prog}

To observe the spatial spread of the reduction of demand for each run,
we group the grid's demand nodes into bins based on their ``hop distance''
from the original failed line. 
In each bin we compute the local yield $Y(t,h)$.  We
average the local yield $Y(t,h)$ for runs resulting in large blackouts
(Fig.~\ref{subfig:CurrCons-USWI}).  The yield in each bin at the end
of cascades resulting in large blackouts is almost independent of the
distance from the initially failed line.  While at the beginning of
the cascade the blackout is localized near the initially failed line,
eventually the blackout spreads uniformly over the entire system.
Delocalization occurs at the end of the latent period of the cascade.
This can be clearly seen from the behavior of the blackout profiles,
which start to rapidly drop down for large distances only at
intermediate time steps of the cascade.

To give a more quantitative measure of the blackout spread, we use the
``blackout radius of gyration'' $\left(r_B(t)\right)$ metric defined in Section
\ref{sec:metrics}.  Fig.~\ref{f:RadGyrMod} shows the behavior of
$r_B(t)^2$ versus the cascade time step $t$ for the cascades which result
in small blackouts (Fig.~\ref{subfig:radGyr-nb-USWI}) and large
blackouts (Fig.~\ref{subfig:radGyr-b-USWI}).  We observe the same
phenomena --- initially $r_B(t)^2$ grows slowly in the runs resulting
in large and small blackouts.  However, while in
runs resulting in small blackouts the cascade stops during this latent
period, in runs resulting in large blackouts the cascade starts to
rapidly spread over a large area.  It should still be noted, though,
that the rate of this spread decreases when tolerance increases.

\subsection{Main Lessons Learned from the Cascades in the USWI model}
\label{sec:LessonsUSWI}

\begin{enumerate}

\item The yield has a pronounced bimodality, for which a grid suffers either a large blackout with $Y<80\%$ or a very small reduction of demand with $Y>90\%$. 
\item Increasing tolerance $\alpha$ decreases the probability $\Pi(\alpha)$ of a
  large cascade with low yield.
\item The higher the significance of the initially failed line $u$,
  the larger the tolerance $\alpha$ necessary to prevent cascades.
  Thus, the same effect can be achieved by protecting a certain
  fraction of the most important lines from spontaneous failure as by
  increasing the tolerance $\alpha$ of all lines.
\item $\langle Y\rangle$ strongly depends on $\alpha$ and weakly
  depends on $u$.  It increases with $p$ for only $p>0.7$.
\item $\langle L\rangle$ also strongly depends on $\alpha$ and weakly
  depends on $u$.  It increases with $p$ even for small $p$.
\item $\langle G\rangle$ weakly depends on $\alpha$ and practically
  does not depend on $p$ and $u$.  This suggests that in large
  blackouts the grid disintegrates into very small clusters.
\item Cascades which do not result in large blackouts are usually
  short.  Cascades resulting in large blackouts usually take a large
  number of time steps, and the number of time steps increases with
  $\alpha$.
\item There is a latent period during which the reduction in demand is
  small and not many lines have failed.  During this period, it may be
  possible to intervene and optimally redistribute current flow to
  prevent the cascade from growing uncontrollable and to prevent a
  large blackout.  The duration of the latent period linearly
  increases with $\alpha$.
\item In cascades resulting in large blackouts, the reduction in
  demand usually begins in the vicinity of the initially failed line and
  the blackout remains localized near the initially failed line until
  the end of the latent period.
\item For cascades resulting in large blackouts, the final local yield is
  almost independent of the distance from the initially failed line.

\end{enumerate}

\section{Degree And Distance Attachment Model}
\label{sec:prefModel}

In the previous section we find that cascading failures in the USWI
model have characteristic features of a first-order transition: the
bimodal distribution of yield and the latent period during which the
damage to the network is insignificant. It is important to investigate
whether these features are due to particular characteristics of the
USWI design, or whether they are universal features of a much broader class of
models. Moreover, the data on real grids are limited and therefore it
is important to develop algorithms for generating synthetic grids
resembling known real grids. The two basic features of USWI that we
would like to reproduce are the degree distribution and the
distribution of line lengths.  The degree distribution of the USWI
discussed in Section \ref{sec:WestCoast} is in agreement with the
Barab\'{a}si-Albert preferential attachment model
\cite{bararev,barasci}. Accordingly we use the Barab\'{a}si-Albert
model as the basis of the synthetic model. In the original
Barab\'{a}si-Albert model, a newly created node is attached to an
existing node with a probability proportional to the degree of this
node. However, for power grids embedded in two-dimensional space, the
length distribution of lines, resulting from the degree preferential attachment,
would not decrease with the
length. Therefore, in order to create a grid with a decreasing
length distribution, one must introduce a penalty for attaching to a
distant node. Here we will employ the Degree And Distance Attachment
(DADA) model with a distance penalty developed in
Refs. \cite{xulvi,manna}.  This method produces degree and length
distributions similar to those of the USWI (see Section
\ref{sec:USWIdata}).

The DADA model randomly generates nodes $j=1,2, ... n$ on a plane with a
uniform density one by one. It connects each new node $j$ to an existing
node $i$ based on $i$'s degree and distance with probability 
${\mathrm P}(\{i,j\})\propto{k_i}/{r_{ij}}^\mu$, where $k_i$ is a current
degree of node $i$ and $r_{ij}$ is the distance between nodes $i$ and
$j$.  This rule mimics the way real networks are evolved.  A real
network such as the USWI is not planned all at once; rather, new
stations are added to the grid as necessity dictates. The probability
of connection ${\rm P}(\{i,j\})\propto{k_i}/{r_{ij}}^\mu$ is assumed
to be proportional to $k_i$, since connections to nodes of high degree
are more reliable, but also inversely proportional to a power of
$r_{ij}$, since construction of long transmission lines costs more.
The distance penalty $\mu$ is a factor which seeks to optimize the
balance between reliability and cost (see Algorithm
\algref{alg:prefConstr}{alg:prob}). As in USWI, we assume that in the
DADA model $R_{ij}=\rho r_{ij}$, where $\rho$ is resistivity, which is
constant for all the lines in the system.

\begin{figure*}[ht]
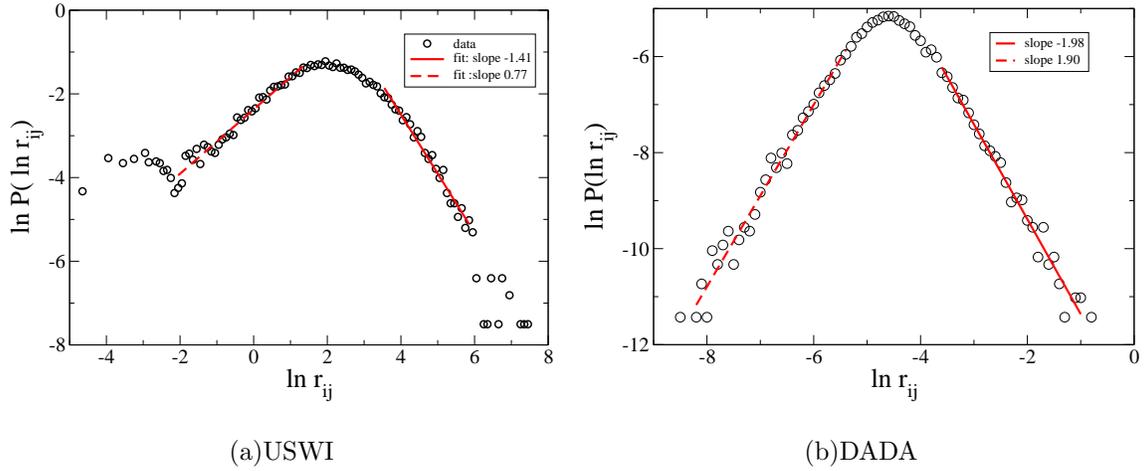

\centering
	\subfigure[USWI]{
  		\vspace*{-2 em}
  		\includegraphics[clip, width=0.44\linewidth,angle=0]{Fig3a.eps}
  		\label{subfig:ResUSWI}
  	}
  	\subfigure[DADA]{
  		\vspace*{2em}
	  	\includegraphics[clip, width=0.45\linewidth,angle=0]{Fig3b.eps}
	  	\label{subfig:ResDADA}
	}
\caption{Distribution of the line lengths, which are the same as resistances, in \subref{subfig:ResUSWI}: the USWI power grid, and \subref{subfig:ResDADA}: the DADA with $\mu=6$, $\ell=1.5$.}
\label{f:res}
\end{figure*}

Refs. \cite{xulvi,manna} show that for $\mu<1$, the degree
distribution of the DADA model is a power law $P(k)\approx k^{-3}$, while
for $\mu>1$, it becomes a stretched exponential
\cite{powerlaw}. However, the fat-tail of the stretched exponential can
be approximated by a power law $P(k)\approx k^{-\gamma}$ with an
exponent $\gamma>3$ (Fig.~\ref{subfig:DegDADA}). Ref. \cite{manna}
also shows that the line length distribution
$P(r_{ij})\approx r_{ij}$ for $r_{ij}\to 0$, and for large $\mu$,
$P(r_{ij})\approx r_{ij}^{-3}$ for $r_{ij}\to \infty$. The functional
form of $P(r_{ij})$ for the DADA and USWI models are similar, but the
exponents are different.  As mentioned in Section
\ref{sec:USWILengthDist} regarding the USWI model, these asymptotic behaviors correspond to the
slopes $\nu_-=2$ and $\nu_+=-2$ of the logarithmic distribution
$P(\ln(r_{ij}))$ observed in the DADA model (Fig.~\ref{subfig:ResDADA}), 
while for the USWI
model these values are $\nu_-=0.77$ and $\nu_+=-1.43$.  In our
simulations, we select $\mu=6$. For this choice of $\mu$, the degree
distribution exponent $\gamma\approx 4.3$, while $-\nu_+=2$. The
corresponding values in the USWI model are smaller. Both $\gamma$ and
$-\nu_+$ can be decreased by decreasing $\mu$, so that the degree and
length distributions of the DADA model would be closer to those of the
USWI. However, upon doing so, our results on the distribution of currents
in the DADA model and properties of the cascading failures do not
change significantly, indicating that the observed features of the cascades are
quite universal. The discrepancy in $\nu_-$ for the DADA and USWI
is related to the fact that in the DADA model the nodes are spread on
the plane with a uniform density, while in the USWI model the density
of nodes is related to the population density which has fractal-like
features.

\subsection{Construction of the DADA model}
\label{sec:prefConstruct}
In this subsection, we will describe in detail the construction of the
DADA model \cite{xulvi,manna}.  We generate the coordinates $(x_i,y_i)$
of $n$ nodes randomly with a homogeneous density $n/S^2$ over the
($S\times S$) square with periodic boundary conditions. For this periodic square, the distance $r_{ij}$ between two points with coordinates
$(x_i,y_i)$ and $(x_j,y_j)$ is computed as \begin{equation}
  r_{ij}=\sqrt{\Delta x^2 +\Delta y^2},
\end{equation}
where $\Delta x=\min\left(|x_i-x_j|, |S-|x_i-x_j||\right)$ and $\Delta
y=\min\left(|y_i-y_j|, |S-|y_i-y_j||\right)$. We select
$S=1$. Periodic boundary conditions are often used in statistical
physics to minimize finite size effects \cite{Rapaport}.  Our goal is
to create a grid with a given number of lines, $l$.
\begin{figure}
\begin{algorithm}[H]
\caption{DADA Model Construction}
\label{alg:prefConstr}
\begin{algorithmic}[1]
\State Select a random subset $\Lambda$ of size $l-\bar{\ell}n$ from
$\{1,2 ... n\}$.
	\For {$j\gets 1$ {\bf to} $n$} 
	\State Choose coordinates $x_j$ and $y_j$ between $0$ and $S$. \label{alg:coord}
	\State Choose integer number $\ell_j$: 	\label{alg:lines}	
	\If{$j\in \Lambda$} 
		\State $\ell_j=\lfloor \bar{\ell} \rfloor+1$.
    \Else 
        \State $\ell_j=\lfloor \bar{\ell} \rfloor$.
    \EndIf
	\If {$j\leq\ell_j$} \label{alg:jleqellj}
	\ForAll{$i<j$}
	\State Connect $j$ to $i$.
	\EndFor
        \State {$k_j\gets j-1$}.
	\Else
        \State {$k_j\gets 0$}.
	\While{$k_j < \ell_j$} \label{alg:rule}
	\ForAll{$i<j$} 
		\If{$j$ is not directly connected to $i$}
			\State Compute $r_{ij}$. \label{alg:dist}
			\State Assign Probability $\mathrm{P}(\{i,j\})\propto{k_i}/{r_{ij}}^\mu$. \label{alg:prob}
		\Else
			\State $\mathrm{P}(\{i,j\})=0$. \label{alg:P0}
		\EndIf
	\EndFor
	\State Choose node $i$ from distribution $\mathrm{P}(\{i,j\})\propto{k_i}/{r_{ij}}^\mu$ and connect it to $j$. \label{alg:connect}
        \State {$k_j \gets k_j+1$}.
	\EndWhile
	\EndIf
        \EndFor
	\State \Return $G$.
\end{algorithmic}
\end{algorithm}
\end{figure}
We start by randomly placing nodes onto the grid one by one until we
have a total of $n$ nodes (see Algorithm
\algref{alg:prefConstr}{alg:coord}).  When each new node is created,
we connect it on average to $\bar{\ell}=l/n=\langle k\rangle/2$
preexisting nodes in order to achieve our goal of creating a grid with
a given number of lines $l$.  Since $\bar{\ell}$ is a real number, we
preassign to each node $i$ an integer $\ell_i$, the number of lines by
which it will be connected to the previously generated nodes,
according to Algorithm \algref{alg:prefConstr}{alg:lines}.  We randomly
select $l-n\lfloor \bar{\ell} \rfloor < n$ nodes, where $\lfloor
\bar{\ell} \rfloor$ is the integer part of $\bar{\ell}$.  For these
nodes, we choose $\ell_i=\lfloor \bar{\ell} \rfloor+1$.  For the rest
of the nodes, we choose $\ell_i=\lfloor \bar{\ell} \rfloor$.  For each
new node $j$, we attempt to create $\ell_j$ lines with the previously
existing nodes.  If $j\leq\ell_j$, then we connect $j$ to all
preexisting nodes (see Algorithm \algref{alg:prefConstr}{alg:jleqellj}), as
we cannot create $\ell_j$ lines without duplicating lines.  If
$j\geq\ell_j$, there are more existing nodes than $\ell_j$ and we
create lines according to the following rule (see Algorithm
\algref{alg:prefConstr}{alg:rule}).  First we compute the distances
$r_{ij}$ between the new node $j$ and all existing nodes $i<j$ (see
Algorithm \algref{alg:prefConstr}{alg:dist}).  Next we assign to each
existing node $i$ a probability $\mathrm{P}(\{i,j\})$ of connecting to
the new node $j$, proportional to ${k_i}/{r_{ij}^\mu}$, where $k_i$ is the
degree of node $i$ and $\mu$ is a parameter giving the penalty for
distance (see Algorithm \algref{alg:prefConstr}{alg:prob}).  We then
connect the new node to a node chosen from this probability
distribution (see Algorithm \algref{alg:prefConstr}{alg:connect}).  We
repeat this step $\ell_{j}$ times, with the additional condition that
all nodes $i$ already directly connected to node $j$ have probability
of connecting $\mathrm{P}(\{i,j\})=0$ instead of
$\mathrm{P}(\{i,j\})\propto{k_i}/{r_{ij}^\mu}$ (see Algorithm
\algref{alg:prefConstr}{alg:P0}).  At the end, a total of almost
$n\ell=l$ lines are created.

For the USWI network $\bar{\ell}=\langle k \rangle/2 \approx 1.5$, so for the DADA model we choose $\bar{\ell}=1.5$ (more accurate values of $\bar{\ell}$ do not significantly affect our results).
As mentioned in Section \ref{sec:WestCoast}, the average degree $\left<k\right>$ of the USWI network is 2.67. For supply nodes, it is slightly larger (2.88); for demand nodes, it is very slightly smaller (2.61).
Averaging over 100 different grids, our DADA model has slightly higher average degree of 2.84. 
More accurate values of $\bar{\ell}$ do not significantly change the cascading properties of the DADA model.

After placing $n$ nodes and $l$ lines, we randomly assign $n^+$ supply nodes and (different) $n^-$ demand nodes.
Our network was simulated with $n=13135$, $n^-=3888$, and $n^+=1197$ to match the USWI. We assign the supply and demand nodes independent of degree. Thus the average degrees of the supply and demand nodes are the same as the average degree of the DADA model.

\begin{figure}[ht]
\centerline{
  \includegraphics[clip, width=0.5\linewidth,angle=0]{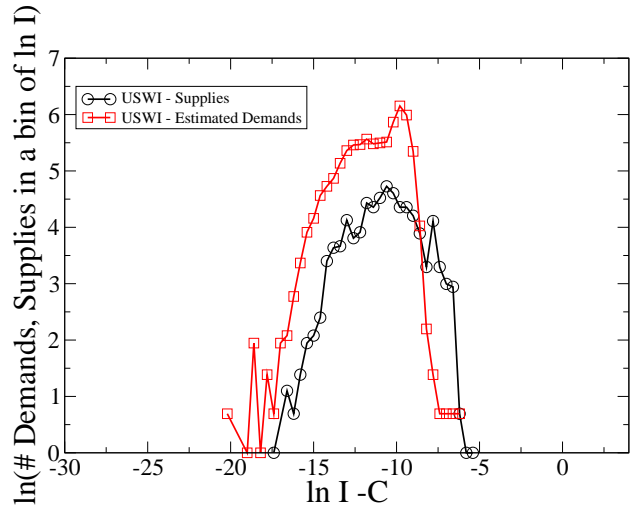}}
\caption{Distribution of the supplied and demanded currents in the USWI power grid. Currents are portrayed in arbitrary units.} 
\label{f:pow}
\end{figure}

Since the supplies and demands of the USWI have an approximately lognormal distribution (see Fig.~\ref{f:pow}), we generate currents of supplies and demands in the DADA model 
following a modified lognormal distribution: 
\begin{equation}
I_i^+ = e^{\nu_i \sigma^++m^+\ln{k_i}},
\end{equation}
for supplies and
\begin{equation}
I_i^- = e^{\nu_i \sigma^-+m^-\ln{k_i}},
\end{equation}
for demands, where $\nu_i$ is randomly generated according to a standard normal distribution, $\sigma^\pm$ is a standard deviation, and $m^\pm$ is a parameter which creates a correlation between the node's current and its degree.

Furthermore, since it is unrealistic to have nodes with very high supply and demand vales, we introduce a cut-off $a^\pm\sigma^\pm$, where $a^\pm$ is a parameter of the model such that we accept only 
$I_i^\pm\leq e^{a^\pm\sigma^\pm}$.
Thus, the supply and demand of each node is
\begin{equation}
I_i^\pm = \min\left(e^{\nu_i \sigma^\pm+m^\pm\ln{k_i}}, e^{a^\pm\sigma^\pm}\right).
\end{equation}
This cut-off corresponds to the sharp drops of the right tails of the supply and demand distributions 
in the USWI (Fig.~\ref{f:pow}).

To best match the USWI data, we let the values of $m^\pm$ be the slopes of the regression lines
of the log-log scatter plots which plot the average supply/demand versus the degree of corresponding nodes.
We then select values of $\sigma$ and $a$ so that the distributions simulated for the DADA model best match the USWI distributions (Fig.~\ref{f:pow}).
We obtain $\sigma^+=2.0$, $m^+=0.38924$, $a^+=1.6$, $\sigma^-=1.8$, $m^-=0.62826$, and $a^-=1.2$.

\section{Comparison of the USWI and DADA Models}
\label{sec:comparison}

\begin{figure}[h]
\centerline{
  \includegraphics[clip, width=0.5\linewidth,angle=0]{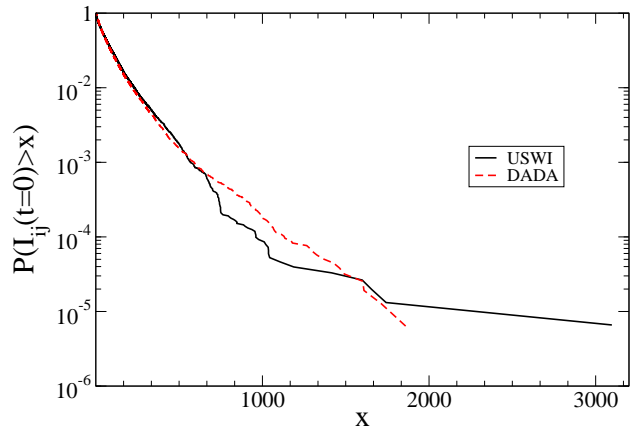}}
\caption{Cumulative distribution of currents for the USWI model power grid and DADA model with $\mu=6$ and $\ell=1.5$. (Currents are measured in arbitrary units.)}
\label{f:curr}
\end{figure}

Here we compare the main results from the USWI model and the DADA model.
We also discuss reasons for the differences observed.
The cumulative distribution of currents in the DADA model closely follows the exponential distribution of currents in the USWI grid (Fig.~\ref{f:curr}). This is important because the ratio of currents in the two 
models corresponding to the same significance of lines $u$ is approximately constant.

\subsection{Evolution of Cascades}
\label{sec:EvolutionComp}

The distribution of the yield $Y$ in the DADA model is also bimodal
for approximately the same set of parameters $\alpha$, $u$, and $p$ as
in the USWI model, but the gap between the two modes is significantly
wider in the DADA model than in the USWI model (Fig.~\ref{f:PIf}).
Figure ~\ref{f:PIf} compares the yield distributions of the USWI and
DADA models for $u=1$ and $p=0.9$ for several values of $1\leq
\alpha<2$.  Both models always collapse ($Y<0.8$) for small values of
$\alpha$, and survive ($Y>0.8$) for large values of $\alpha$. But for
the DADA values, chances of large blackout (risk) are smaller for the
same set of parameters than in the USWI model. For example, the DADA
model can still survive with a small probability for $\alpha=1.2$, but the
USWI always collapses for $\alpha <1.3$. Conversely, we do not observe
any large blackouts in the DADA model for $\alpha>1.7$, while the USWI
model can still have large blackouts even for $\alpha=1.9$.  Thus,
even though in the event of a large blackout the average yield in the
USWI is greater than in the DADA model (and thus, in this sense, the
DADA model is more vulnerable than the USWI model), the risk of large
blackouts is greater in the USWI model than in the DADA model for the
same set of parameters.

These differences may be related to the fractal structure of the USWI, in which
densely populated areas with lot of demand and supply nodes are
separated by large patches of empty land over which few long
transmission lines are built, whereas the DADA model
has constant density of nodes.  Thus, it is less likely that the
cascade will spread over the entire grid in the USWI model than in the
DADA model, but a higher tolerance is necessary to prevent large
blackouts in the USWI model than in the DADA model.

Qualitatively, the behaviors of the metrics $\langle Y\rangle$,
$\langle L\rangle$, and $\langle G\rangle$ are similar in the USWI
model and in the DADA model, but in the DADA model the survival
quantities are always smaller for the same $\alpha$, $u$, and $p$.
This indicates that, remarkably, the artificial DADA model is more
vulnerable than the USWI model based on actual data
(Figs.~\ref{f:risk}, \ref{f:alpha0}, and \ref{f:GYL}).  The values of
$\left<G\right>$ in the DADA model are very small, indicating that in
the event of a large blackout the DADA network disintegrates into very
small connected components, each constituting about 1\% of the nodes
of the grid. In the USWI grid, the average largest component is larger,
because USWI grid consists of several dense areas connected by few
long lines. The overload of these long lines breaks the USWI grid into
relatively large disconnected components, preventing the cascade from
further spreading.

\subsection{Cascade Temporal Dynamics}
\label{sec:temporalComp}

\begin{figure*}[h]
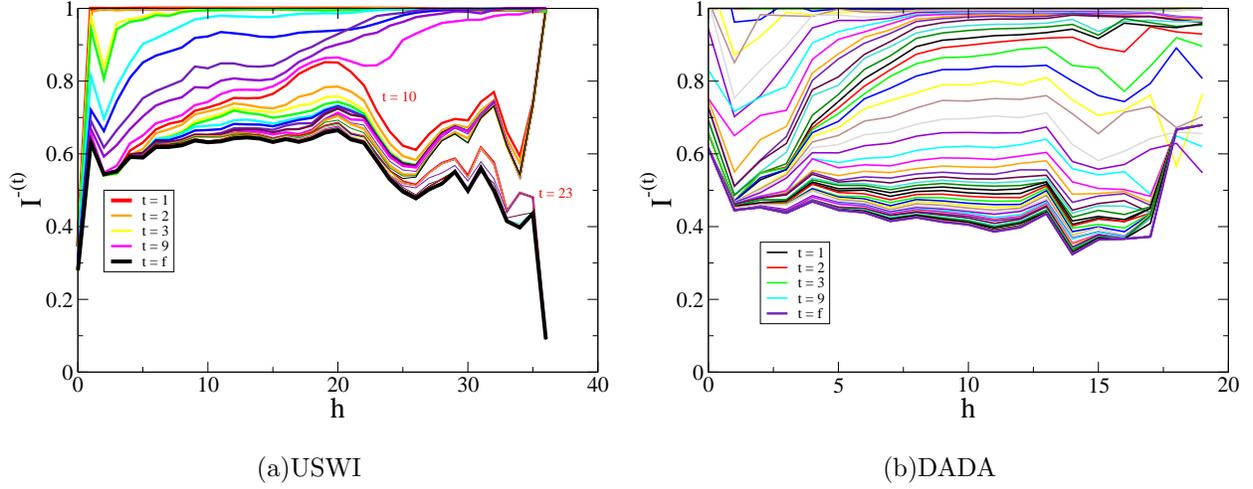

\centerline{
  \subfigure[USWI]{
  		\includegraphics[clip, width=8cm,angle=0]{Fig13a.eps}
  		\label{subfig:CurrCons-USWI}
  }
  \subfigure[DADA]{
  		\includegraphics[clip, width=8cm,angle=0]{Fig13b.eps}
  		\label{subfig:CurrCons-DADA}
  }}
\caption{The fraction of current reaching demand nodes as a function of hop distance in runs resulting in large blackouts for different cascade time steps, with $p=0.9$ and $\alpha=1.6$, \subref{subfig:CurrCons-USWI} for the USWI data, and \subref{subfig:CurrCons-DADA} for the DADA model with $\mu=6$, $\ell=1.5$.}
\label{f:CurrByDist}
\end{figure*}

The spatial and temporal behaviors of the cascades in the DADA model
closely follow the behaviors in the USWI model (Figs.~\ref{f:Ncasc},
\ref{f:latent}, \ref{f:CurrByDist}, and
\ref{f:RadGyrMod}).  $r_B(t)^2$ in the DADA model is much smaller than in
the USWI model due to the different structures of the models and
difference in diameters of the networks.  The longest distance (in
terms of number of hops) between any two nodes (i.e. diameter of the
network) in the DADA model is $\approx 16$, while in the USWI model it is
$\approx 41$.  In both models we see that the cascade spreads more quickly
for smaller $\alpha$ than for larger $\alpha$.
However, the first-order all-or-nothing nature of the cascades,
characterized by a latent period during which the blackout is small
and localized followed by a fast blackout spread over a large area, is
common in both models.

\subsection{Cascade Spatial Spreading}
\label{sec:spatialComp}

\begin{figure*}[h]
\centerline{
  \includegraphics[clip, width=\linewidth,angle=0]{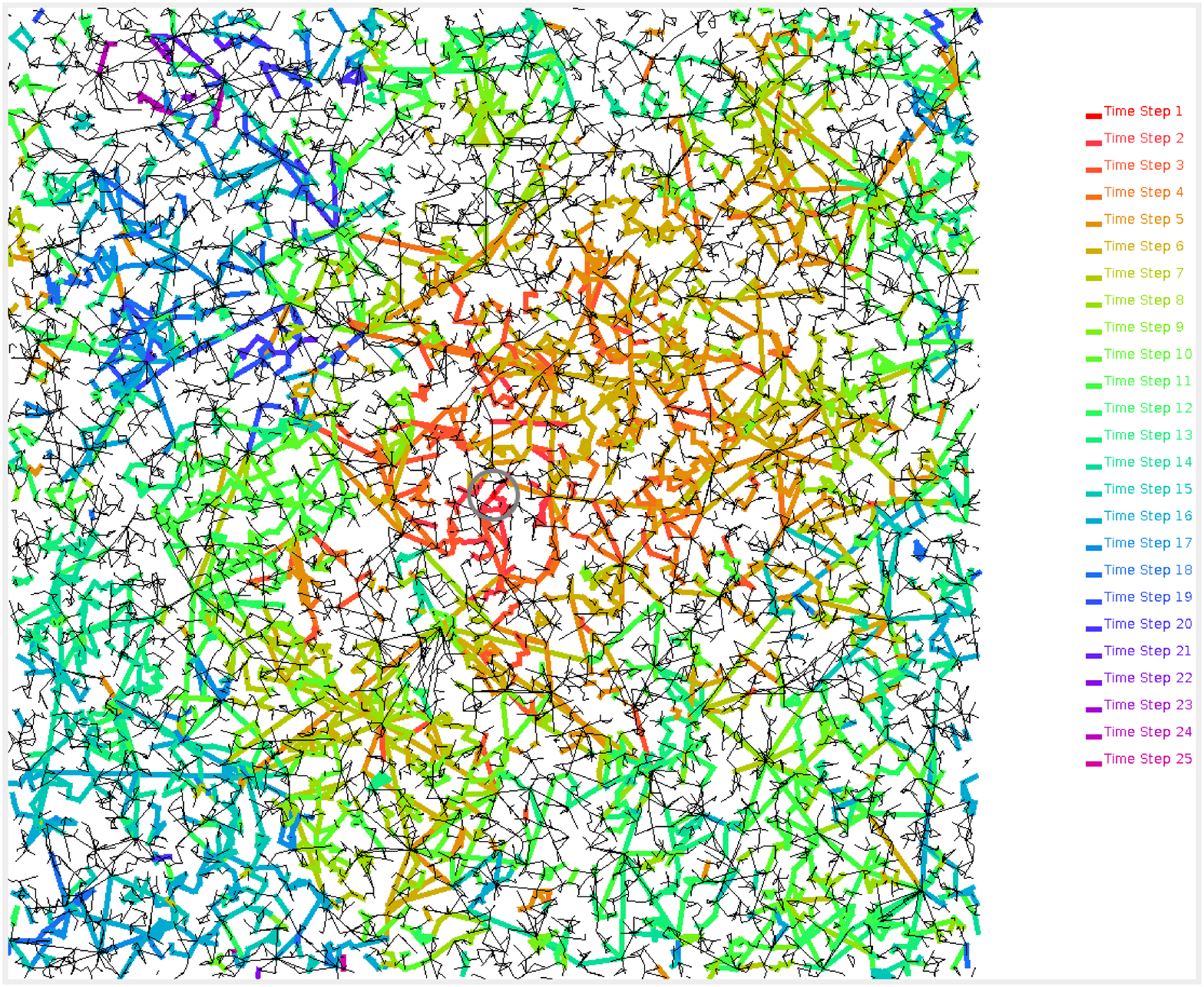}}
\caption{(Color online) Propagation of a cascade of failures for $\alpha=1.8$, $p=0.4$, $u=1$ in the DADA model with 13135 nodes, $\ell=1.5$, $\mu=6$. Lines that failed at different time steps of the cascade are shown with different colors. The line randomly selected to fail due to spontaneous failure or attack is portrayed as the center of the grid and is surrounded by a gray circle.}
\label{f:grid}
\end{figure*}

The DADA model has an advantage over the USWI model, that in the
former we know the exact coordinates of the nodes and thus we can
illustrate the spatial and temporal evolution of the blackout as a
sequence of snapshots on the $(x,y)$ plane.  Fig.~\ref{f:grid} shows
spatial snapshots of the cascading failures taken at different time
steps for the DADA model with parameters $\alpha=1.8$ and $p=0.4$.
The color of a line indicates the time step of the cascade at which
this line has failed.  One can see that during first 3 time steps of
the cascade (red lines) the area of line failures is small and
localized near the initial failure.  The cascade starts to spread
during time steps 4-8 (orange and yellow-green), but the area of line
failures is still localized.  At time step 10 the cascade quickly
spreads to very distant parts of the system (green).  The blue and
violet lines are the final time steps of the cascade.  Thus, the
figure suggests that there is a latent period of the cascade during
which the area of line failures is small and localized.

\section{Conclusion}
\label{sec:conc}

The DADA model and the USWI model have many common features.  The
physical features, such as the distribution of degrees, resistances, and
currents, compare well in both models.  The behavior of the cascades
of failures in the DADA model is also similar to their behavior in the
USWI model power grid, despite the differences in 
construction of these models.

The overloads and cascading failures in the USWI and DADA models have
features of all-or-nothing transition, just like in a broad spectrum
of more primitive models such as the Motter
model \cite{MotterLai,Motter}.  In the Motter model, instead of
currents, the betweenness of each node in a graph is computed and the
maximum load of each node is defined as its original betweenness multiplied by the
tolerance.  Then, a random node is taken out, simulating an initial
failure, and the new betweenness of each node is calculated. If the new
betweenness of a node exceeds its maximum load, this node is taken out
and the entire process is repeated. The yield in the Motter model is
defined as the fraction of survived nodes at the end of the cascade. The
distribution of the yield in the Motter model is bimodal for a large
range of tolerances.  Similarly, in a wide range of parameters, both
the DADA model and the USWI power grid are in a metastable state and
there exists the risk that the failure of a single line will lead to a
large blackout, in which the yield falls below 80\%. As tolerance
increases beyond $2.0$, the risk of a large blackout decreases almost
to 0.

The level of line protection, $p$, increases the robustness of the
grid, but to a lesser extent than does the tolerance.  An important
parameter defining the robustness of the grid is the significance of
the initial failure $u$. The risk of a large blackout increases with
$u$ in the same way as with tolerance.  Given a particular $\alpha$,
when $u$ is small, there is practically no risk of a large blackout,
while when $u$ approaches 1, the risk is maximal for the given
$\alpha$.  If $\alpha$ is kept constant and $u$ decreases, there is the
same effect on the risk of a large blackout as when $\alpha$ increases
and $u$ is kept constant, meaning that the same effect could be
achieved by protecting important lines as by increasing the overall
tolerance. Nevertheless, even if the initially failed lines are
selected irrespective of their currents ($u=\Delta u=1$), the
distribution of yields remains bimodal, but the probability of large
blackouts significantly decreases.
 
Upon failure of a line, the first few cascade time steps affect only
the immediate vicinity of the failed line.  This is the latent period
of the cascade, during which it may be feasible to intervene and
redistribute the current flow and possibly prevent a large
blackout from occurring.  At some time step, the cascade may begin to spread
quickly, and only then will a large blackout occur.  In this case, the
blackout radius increases rapidly with cascade time steps, but with a
lower rate for higher values of $\alpha$.  Once the failure has spread
over the entire grid, the cascade continues to overload a small number of
lines before terminating.  At this time step the grid completely
disintegrates into small disconnected clusters.  During the first few
time steps of the cascade, the demand does not significantly
decrease, but starts to quickly drop at the end of the latent period.
The duration of the latent period of the cascade linearly increases
with the overall tolerance and provides sufficient time for grid
operators to intervene and stop the cascade.

Similar phenomena were observed during the large blackout of August
14th 2003.  It is well documented \cite{NYISO} that the power outage,
which affected a large portion of the Northeastern region of the USA
on August 14th 2003, was caused by line overloads due to a heat-wave.
But what might have been a manageable local blackout cascaded into
massive widespread distress on the electric power grid.  Power was not
re-distributed after overloaded transmission lines hit unpruned
foliage, and the spreading of blackout was exacerbated over a time
period of 30 minutes.
This example closely resembles the large blackout scenario predicted by
our models. Thus, our model provides useful understanding of general features of
the cascades of failures in power grids, which may be used for
increasing the resilience of power grids and designing optimal
shedding strategies for preventing cascades from spreading.

\section{Acknowledgments}
\label{sec:ack}
This work was supported in part by DTRA grants HDTRA1-10-1-0014, HDTRA1-14-1-0017, and HDTRA1-13-1-0021, DARPA RADICS under contract \#FA-8750-16-C-0054, DOE GMLC program, and the People Programme (Marie Curie Actions) of the European Unions Seventh Framework Programme (FP7/2007-2013) under REA grant agreement no. [PIIF-GA-2013-629740].11.
We also acknowledge the partial support of this research through the Dr. Bernard W. Gamson Computational Science Center at Yeshiva College.
We thank Meric Uzunoglu and Andrey Bernstein for their help with processing the USWI data. 
We also thank Guifeng Su for his work on programming the preferential attachment algorithm and measuring ``hop distance'', Yehuda Stuchins for his preliminary work on studying relations between distance and failures, and Tzvi Bennoff for testing alternative parameters. 
We appreciate the assistance of Adam Edelstein, Jonathan Jaroslawicz, and Brandon Bier in sorting the data. 
We are grateful to S. Havlin, G. Paul, and H.E. Stanley for productive interactions.

\clearpage
\section{Figures}
\label{sec:figs}

\begin{figure*}[h]
\centering
	\subfigure[ USWI, $p=0.5$]{
  		\includegraphics[clip, width=0.45\linewidth,angle=0]{Fig8a.eps}
  		\label{subfig:YvA-p0.5-USWI}
  	}
  	\subfigure[ USWI, $u=0.9$]{
  		\includegraphics[clip, width=0.45\linewidth,angle=0]{Fig8b.eps}
  		\label{subfig:YvA-u0.9-USWI}
  	}\\
	\subfigure[ DADA, $p=0.5$]{
		\includegraphics[clip, width=0.45\linewidth,angle=0]{Fig8c.eps}
		\label{subfig:YvA-p0.5-DADA}
	}
	\subfigure[ DADA, $u=0.9$]{
		\includegraphics[clip, width=0.45\linewidth,angle=0]{Fig8d.eps}
		\label{subfig:YvA-u0.9-DADA}
		}
\caption{Probability of large blackout $P(Y<0.8)$, or risk $\Pi(\alpha)$, as function of $\alpha$. \subref{subfig:YvA-p0.5-USWI} USWI model
for different values of $u$ and $p=0.5$. \subref{subfig:YvA-u0.9-USWI} USWI model for different values of $p$ and $u=0.9$. \subref{subfig:YvA-p0.5-DADA} DADA model for different values of $u$ and $p=0.5$. \subref{subfig:YvA-u0.9-DADA} DADA model for different values of $p$ and $u=0.9$.}
\label{f:risk}
\end{figure*}

\begin{figure*}[h]
\centerline{
  \subfigure[USWI, $p=0.5$]{
  		\includegraphics[clip, width=0.45\linewidth,angle=0]{Fig11a.eps}
  		\label{subfig:tVsAlph-USWI-p0.5}
  }
  \subfigure[USWI, $u=0.9$]{
  		\includegraphics[clip, width=0.45\linewidth,angle=0]{Fig11b.eps}
  		\label{subfig:tVsAlph-USWI-u0.9}
  }}
\centerline{
  \subfigure[DADA, $p=0.5$]{
  		\includegraphics[clip, width=0.45\linewidth,angle=0]{Fig11c.eps}
  		\label{subfig:tVsAlph-DADA-p0.5}
  }
  \subfigure[DADA, $u=0.9$]{
  		\includegraphics[clip, width=0.45\linewidth,angle=0]{Fig11d.eps}
  		\label{subfig:tVsAlph-DADA-u0.9}
  }}  
\caption{Dependence of the average duration of the cascade on tolerance $\alpha$. \subref{subfig:tVsAlph-USWI-p0.5} USWI model and \subref{subfig:tVsAlph-DADA-p0.5} DADA model for different $u$ at $p=0.5$. \subref{subfig:tVsAlph-USWI-u0.9} USWI model and \subref{subfig:tVsAlph-DADA-u0.9} DADA model for different $p$ at $u=0.9$. 
The dependence of the duration of the cascade on both $u$ and $p$ is weak. For each $u$ and $p$ there
are two lines: one for cascades resulting in large blackouts with yield $Y<0.8$, and another for cascades resulting
in small blackouts $Y>0.8$. The duration of the cascades in the large blackout cases is always larger than 10 and is increasing
with $\alpha$, while for the cases with small blackouts the cascades are short. Here we see the cascade slows down with higher $\alpha$, so higher $\alpha$ means higher resilience and longer latent periods. This shows the resilience of the grid is based on $\alpha$.}
\label{f:Ncasc}
\end{figure*}

\begin{figure*}[h]
\centerline{
  \subfigure[USWI, $Y>0.8$]{
  		\includegraphics[clip, width=0.45\linewidth,angle=0]{Fig14a.eps}
  		\label{subfig:radGyr-nb-USWI}
  	}
  \subfigure[USWI, $Y<0.8$]{
  		\includegraphics[clip, width=0.45\linewidth,angle=0]{Fig14b.eps}
  		\label{subfig:radGyr-b-USWI}
  	}}
\centerline{
  \subfigure[DADA, $Y>0.8$]{
  		\includegraphics[clip, width=0.45\linewidth,angle=0]{Fig14c.eps}
  		\label{subfig:radGyr-nb-DADA}
  	}
  \subfigure[DADA, $Y<0.8$]{
  		\includegraphics[clip, width=0.45\linewidth,angle=0]{Fig14d.eps}
  		\label{subfig:radGyr-b-DADA}
  	}}
\caption{The averaged behavior of the radius of gyration of the cascading  failures
in small blackouts $Y>0.80$ in \subref{subfig:radGyr-nb-USWI} the USWI model and \subref{subfig:radGyr-nb-DADA} the DADA model, and large blackouts with $Y<0.80$ in \subref{subfig:radGyr-b-USWI} the USWI model and \subref{subfig:radGyr-b-DADA} the DADA model.
$r_B(t)^2$ in the DADA model is much smaller than it is in the USWI model due to the different structures of the models and difference in diameters of the networks. The longest distance between any two nodes (i.e. diameter of the network) in the DADA model is $\approx 16$, while in the USWI model it is $\approx 41$.
In both models we see that the cascade spreads more quickly for smaller $\alpha$ than for larger $\alpha$.}
\label{f:RadGyrMod}
\end{figure*}

\clearpage

\appendix

\section{Solving Flow Equqations}
\label{sec:makematrix}

In the direct current approximation, the outgoing current from node $i$ through a transmitting line with resistance $R_{ij}$ is $I_{ij}=(V_i-V_j)/R_{ij}$, where $V_i$ and $V_j$ are the potentials of nodes $i$ and $j$ respectively.
In addition to the transmitting lines we have to take into account the currents generated by the supply nodes $I_{i}^+$ and the currents consumed by the demand nodes $I_{i}^-$.
In the direct current approximation, it is assumed that the currents $I_{i}^+>0$ and $I_{i}^->0$ are equal to the powers $W_{i}^+$ and $W_{i}^-$ generated by the supply nodes and consumed by the demand nodes, respectively, as explained below.
We assume that each supply node $i$ is connected to the source of voltage $V_{i}^+$ by a line with high resistance $R_{i}^+$ and each demand node $i$ is connected to the ground $V_{i}^-$ by a line with high resistance $R_{i}^-$.
For simplicity we assume that $V_{i}^+=V$, where $V$ is a constant, and $V_{i}^-=0$. 

Since the sum of all supply powers must be equal to the sum of all demand powers and resistances $R_{i}^\pm$ are much larger than $R_{jk}$ of any transmission line, $V_i\approx V/2$. 
Therefore we can select $R_i^\pm=V/(2W_i^\pm)$ so that the current $I_i$ flowing through each resistance $R_i^\pm$ is approximately equal to the power $W_i^\pm$. 
To ensure that $R_i^\pm >> R_{jk}$, we select $V=M {\rm max}_i \left(W_i^\pm\right) {\rm max}_{jk}\left(R_{jk}\right)$, where $M$ is a large constant. 
If $M$ increases, our approximation of the power as current  improves in terms of $|I_i-W_i^\pm|$, but the convergence of our algorithm for solving the Kirchhoff equations slows down. 
We find for $M\approx 10^3$, both accuracy and speed are acceptable. 

Thus, the system of the Kirchhoff equations for the grid consists of
$n$ linear equations for the potentials of each node $V_i$
\begin{equation}
V_i\left(\sum_{j \in N(i)}\frac{1}{R_{ij}}\right)-\sum_{j \in N(i)}\frac{V_j}{R_{ij}} =\delta_{i}^+I_{i}^+-\delta_{i}^-I_{i}^-, 
\label{e:Kirch0}
\end{equation}
where $\delta_{i}^+=1$ or $\delta_{i}^-=1$ if a node $i$ is a supply node or a
demand node, respectively. Otherwise, $\delta_{i}^+=\delta_{i}^-=0$.

The determinant of this system is equal to zero, hence solving such a system requires complex linear algebra procedures.
In order to be able to use simple relaxation procedures,
we must regularize this system by adding positive diagonal terms.
We can achieve this by assuming that all the supply nodes are connected to high voltage $V$
with resistance $R_{i}^+=M/I_{i}^+$
and that all the demand nodes are connected to the ground with resistance $R_{i}^-=M/I_{i}^-$,
where $M$ is a constant.
Since units of currents are arbitrary, we decrease $I_{i}^+$ and
$I_{i}^-$ by a constant factor of one million; since $M$ remains constant, this ensures that
$R_{i}^+$ and $R_{i}^-$ are much larger than any transmission line resistance $R_{ij}$, so the potential differences within the grid are much smaller than the potential differences between the grid and the source, and between the grid and the ground. Thus, when considering the grid's relation to these external potentials,
we can assume all the nodes have approximately the same potential
\begin{equation}
V_g=V\frac{\frac{1}{\sum_i\frac{1}{R_{i}^-}}}{\frac{1}{\sum_i\frac{1}{R_{i}^+}}+
\frac{1}{\sum_i\frac{1}{R_{i}^-}}}.
\end{equation}
Since due to conservation of charge
\begin{equation}
M\sum_i\frac{1}{R_{i}^-}=\sum_iI_{i}^-=\sum_iI_{i}^+=M\sum_i\frac{1}{R_{i}^+},
\end{equation}
$V_g=V/2$.
Furthermore, $I_{i}^+R_{i}^+=M = I_{i}^- R_{i}^-$ equals the potential difference between the demand nodes and the ground. Hence,
$M = V_g = V/2$.
Accordingly, Eq.~(\ref{e:Kirch0}) can be rewritten as
\begin{equation}
V_i\left(\sum_{j \in N(i)}\frac{1}{R_{ij}} +\delta_{i}^+\frac{1}{R_{i}^+}+\delta_{i}^-\frac{1}{R_{i}^-}\right)-\sum_{j \in N(i)}\frac{V_j}{R_{ij}} =\delta_{i}^+\frac{V}{R_{i}^+}.
\label{e:Kirch1}
\end{equation}

Now the determinant of this system is not equal to zero due to the presence
of $1/R_{i}^+$ and $1/R_{i}^-$ terms in the diagonal elements of the system.
System (\ref{e:Kirch1}) can be rewritten as
 \begin{equation}
(\mathbf{I}-\mathbf{B})\vec{V}=\vec{V}_0,
\label{e:Kirch2}
\end{equation}
where $\mathbf{I}$ is an identity matrix,
\begin{equation}
b_{ij}=\left\{
\begin{array}{lr}
	0 & for\ j \notin N(i) \\
 \frac{1}{R_{ij}\left(\sum_{j \in N(i)}\frac{1}{R_{ij}} +\delta_{i}^+\frac{1}{R_{i}^+}+\delta_{i}^-\frac{1}{R_{i}^-}\right)} & for\ j \in N(i)
\end{array}
\right.,
\end{equation}
and
 \begin{equation}
(\vec{V_0})_i=\delta_{i}^+\frac{V_{i}^+}{R_{i}^+\left(\sum_{j \in N(i)}\frac{1}{R_{ij}} +\delta_{i}^+\frac{1}{R_{i}^+}+\delta_{i}^-\frac{1}{R_{i}^-}\right)}.
\end{equation}
Since the determinant $|\det \mathbf{B}|<1$, Eq.~(\ref{e:Kirch2}) can be solved iteratively:
\begin{equation}
\vec{V}_{m+1}=\mathbf{B}\vec{V}_m+\vec{V}_0
\end{equation}
with the initial condition $V_i=V/2$.
Solving these equations, we find voltages $(V)_i$ of each node and the current flow through each line $I_{ij}=|(V_i-V_j)/R_{ij}|$.

Notes:\\ 
\begin{itemize}
	\item[$\blacksquare$] To remove a line from the grid, we set its resistance $R_{ij}=\infty$ (i.e., set the conductance to zero to prevent current from traveling through the line, thus the line has been effectively removed).  
	\item[$\blacksquare$] We assume that the tolerance of the lines connecting the supply nodes to $V$ and demand nodes to ground is $\alpha^+=\alpha^-=\infty$, so these lines never fail. 
	Such a model with finite $\alpha^+$ is much more vulnerable to developing large blackouts than the most conservative variant of shedding that we study.
\end{itemize}

\section{Expediting Computation}
\label{sec:expediting}

To speed up the iterative algorithm for solving linear equations,
we treat separately each disconnected cluster.
We pull from the main grid all the relevant information about each
cluster and treat them as grids of their own.
We cut from these clusters
the ``dangling ends'', defined (as in \cite{Coniglio}) as areas of
transmitting nodes connected to the rest of the cluster through
one single node. 
As these ``dangling ends'' have only transmitting nodes and only one point of connection to the rest of the grid, they have zero current in all their lines.
We identify dangling ends by the Hopcroft-Tarjan algorithm
for finding biconnected components
\cite{Hopcroft}. We then solve the system
in Eq.~(\ref{e:Kirch2}) for each smaller cluster individually
and incorporate the data back into the main grid.



\end{document}